\documentclass[11pt]{article}
\usepackage{tmlpreprint}

\usepackage{cite}
\usepackage{amsmath,amssymb,amsfonts}
\usepackage{textcomp}
\usepackage{multirow}
\usepackage{url}
\usepackage{array}
\usepackage{tabularx}
\usepackage{booktabs}
\usepackage{adjustbox}
\usepackage{xurl}

\providecommand{\rev}[1]{#1}
\providecommand{\revb}[1]{#1}

\lablogo{\includegraphics[width=38mm]{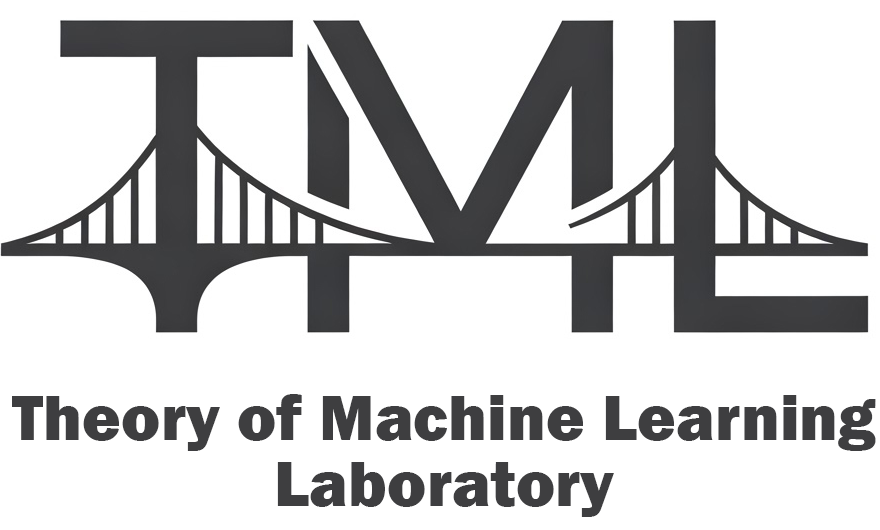}}
\runninghead{Dialogue-Based TTS Lessons Generated with LLMs}

\title{A Semi-Automated System for Generating Dialogue-Based TTS Lessons Using Large Language Models: An Exploratory Study of Educational Potential}
\author{Gendo Kumoi\affmark{1}, Fumie Watanabe\affmark{1}, Tota Suko\affmark{2}, Takashi Ishida\affmark{3}, Yuko Kuma\affmark{4}, Manabu Kobayashi\affmark{2}, and Shigeichi Hirasawa\affmark{2}}
\affiliations{\affmark{1}Nagaoka University of Technology, Nagaoka, Niigata, Japan \quad \affmark{2}Waseda University, Shinjuku, Tokyo, Japan \quad \affmark{3}Takasaki City University of Economics, Takasaki, Gunma, Japan \quad \affmark{4}Kyushu Sangyo University, Fukuoka, Fukuoka, Japan}
\correspondence{Corresponding author: Gendo Kumoi (e-mail: kumoi@vos.nagaokaut.ac.jp). F. Watanabe was with Nagaoka University of Technology, Nagaoka, Niigata, Japan, where this work was carried out. She is now with the Center for Data Science, Waseda University, Shinjuku, Tokyo, Japan.}
\preprintnote{Accepted version of the article published in IEEE Access, DOI: \href{https://doi.org/10.1109/ACCESS.2026.3756639}{10.1109/ACCESS.2026.3756639}. Licensed under CC BY 4.0. This work was supported by JSPS Program for Forming Japan's Peak Research Universities J PEAKS Grant Number JPJS00420240017.}

\begin{document}
\maketitle

\begin{abstract}
This study proposes a semi-automated system for generating dialogue-based lessons using Large Language Models (LLMs) and Text-to-Speech (TTS) technology, and examines its educational potential through a practical quasi-experiment. The system is designed to augment rather than replace educators, and adopts a three-stage human-in-the-loop workflow comprising LLM-based generation of slides and narration scripts, educator review, and automated audiovisual integration. Within this workflow we developed a method for automatically generating Expert--Novice dialogue narration informed by cognitive apprenticeship theory.
Two lessons produced by the system were delivered to 245 first-year high school students, alongside one narrated by a recorded instructor voice. Three findings emerged. First, when learners retrospectively compared the formats, none of four evaluation metrics distinguished synthesized narration from the recorded human voice, and equivalence testing supported this. \revb{Second, an exploratory comparison of the two synthesized sessions, whose respondents likely overlap but could not be individually matched, found no general advantage for the dialogue session over the single-speaker session: of twelve primary outcomes, nine showed no difference, while the dialogue session was rated higher on learners' confidence in their own understanding and lower on audio quality, with higher extraneous cognitive load.} Free-text responses attribute the audio cost to voice variation between generation calls rather than to the dialogue format itself. Third, the retained generation records allowed the educator review step to be characterized, yielding seven review criteria and an instance in which an internal contradiction in generated material was corrected before delivery.
Lesson format, content, order, and session length vary together in this design, and all outcomes are learner self-reports\revb{ with no objective measure of learning}. The results therefore describe what was observed rather than isolating the effect of lesson format.
\end{abstract}

\begin{keywords}
Large Language Models, Text-to-Speech, Dialogue-based lessons, Human-in-the-loop, ARCS model, Cognitive Load Theory, Learning engagement, AI-assisted education
\end{keywords}

\section{Introduction}

\subsection{Background and Motivation}

In the field of education, the rapid advancement of Generative AI and Text-to-Speech (TTS) technology is bringing revolutionary changes to the production and delivery of educational content\cite{ref:ai-education, ref:tts-stt-education}.
In particular, educational models such as the Flipped Classroom rely heavily on high-quality pre-class videos\cite{ref:Das2019Flipped, ref:Jensen2018Investigating}, yet producing them requires considerable time and specialized skill, creating a high barrier to entry for many educators\cite{ref:wang2017}.

In recent years, fully automated content-generation solutions have emerged, but they often lack the pedagogical nuance and factual accuracy essential for effective instruction, leaving practical challenges unresolved\cite{ref:gupta2024}.
Moreover, LLM-generated content carries the risk of ``hallucinations''\cite{ref:llm-hallucination}, making human oversight indispensable in educational contexts.

\subsection{Objectives and Contributions}

The objective of this study is to address these challenges by proposing a novel semi-automated generation system that leverages LLMs and TTS technology with the aim of augmenting rather than replacing educators, and to explore its educational potential as reported by learners in an actual classroom.
In system design, we adopted a human-in-the-loop workflow architecture that balances automation with pedagogical control. In particular, we establish a novel methodology for automatically generating Expert-$\times$-Novice dialogue-format narration scripts, inspired by cognitive apprenticeship theory, from a TTS-optimization perspective.
For the evaluation, we examine the utility of the system and learners' reported experience of dialogue TTS lessons through a multi-faceted questionnaire study grounded in the \rev{ARCS (Attention, Relevance, Confidence, Satisfaction) model}, Cognitive Load Theory, and learning engagement theory. \revb{No objective measure of learning was collected, so the evaluation concerns how the lessons were experienced rather than what was learned.}

\begin{figure}[htbp]
  \centering
  \includegraphics[width=\textwidth]{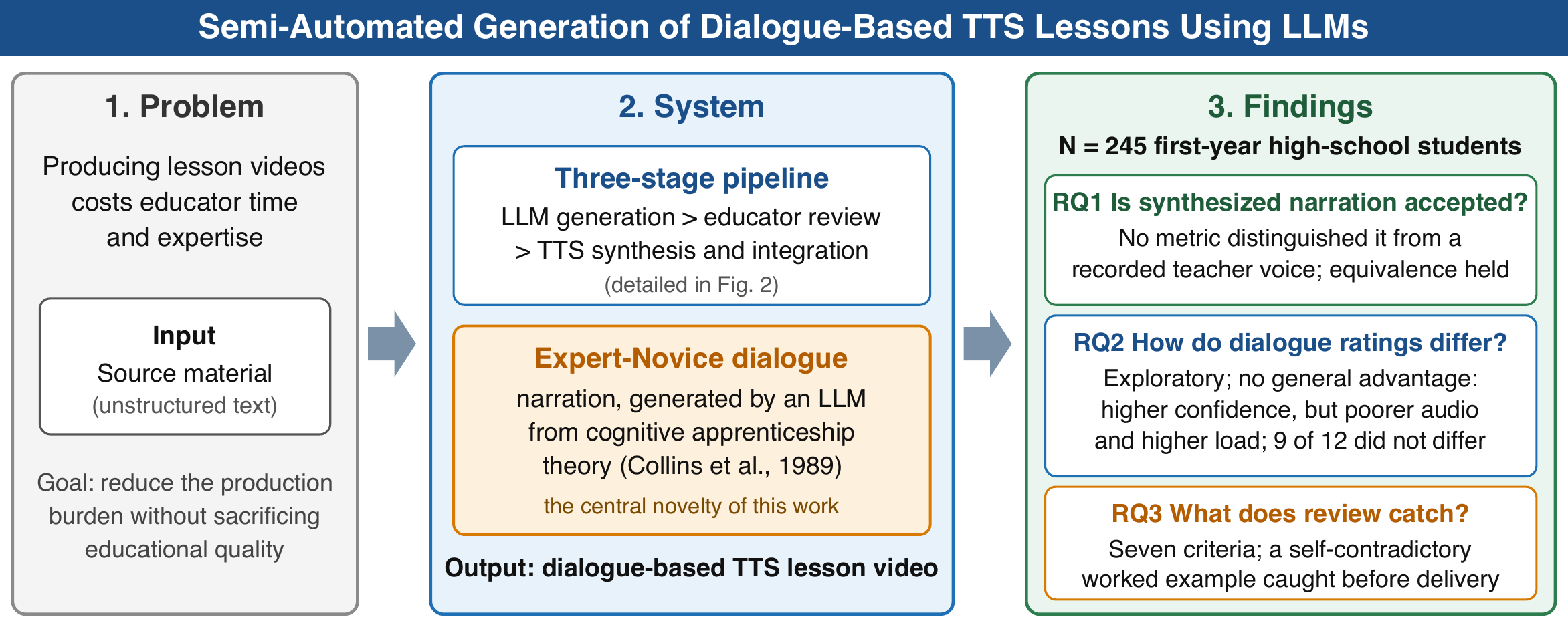}
  \caption{Overview of this study: a three-stage human-in-the-loop system for generating dialogue-based TTS lessons, and what its evaluation found\label{fig:graphical-abstract}}
\end{figure}

\rev{The contribution claimed here is integrative and empirical rather than methodological. We do not propose a new language model or a new speech synthesis method, and the components we use are existing ones. What is new is the combination: a generation pipeline in which the narration is structured by an instructional theory rather than by summarization objectives, an account of the educator review step derived from the complete generation records rather than from description of intent, and an evaluation with 245 secondary school learners in an actual curriculum rather than in a laboratory setting. Prior work on automated slide generation optimizes the fidelity of the generated artifact to a source document; the question taken up here is what happens when such an artifact is narrated, delivered to learners, and evaluated against the alternative those learners would otherwise have received.}
An overview of this study is shown in Figure~\ref{fig:graphical-abstract}.

\section{Related Work}

\subsection{Educational Applications of TTS Technology}

Research on the application of TTS technology to education has been conducted primarily in the contexts of language learning and support for visually impaired learners\cite{ref:tts-stt-education}.
El Fakir et al.\cite{ref:tts-stt-education} demonstrated that TTS and \rev{speech-to-text (STT)} technologies are effective in supporting listening comprehension and pronunciation practice, and they are considered particularly important from the perspective of improving accessibility.

\rev{Closest to the present study, Deng et al.}\cite{ref:deng2026}\rev{ crossed instructor type (a real person against an AI-generated one) with voice type (human against synthetic) in a randomized between-subjects experiment with 126 undergraduates, measuring perceived human-likeness, motivation, and knowledge retention. Voice type had no direct effect on any of the three. What mattered was the instructor: AI-generated instructors were perceived as less human-like, which reduced motivation and in turn retention. Their design is cleaner than ours---content and duration are held constant and assignment is randomized---and it carries an objective retention measure that the present study lacks. It differs in that their materials show an instructor on screen, so the human-likeness pathway they identify runs through the visible persona; the lessons evaluated here consist of slides and narration with no instructor image, leaving voice as the only cue to a human presence.}
In recent years, advances in neural TTS technology have enabled more natural and expressive speech synthesis\cite{ref:neural-tts}, broadening its application to general educational content\cite{ref:dai2022}.

The latest TTS APIs offer, in addition to audio naturalness, the ability to control prosody (rhythm and intonation), pace, and emotional tone via natural-language prompts\cite{ref:gemini-tts}.
This has made it possible to adjust vocal expression according to educational content.

\subsection{Learning Effects of Dialogue-Based Instruction}

Multiple theoretical foundations support the effectiveness of dialogue-based learning.

First, research on vicarious learning has observed that learners who observe dialogues between others and a teacher can reach deeper understanding than through self-study\cite{ref:chi-vicarious}.
In particular, observing a novice pose questions to an expert is thought to help learners clarify their own latent questions and is associated with improved comprehension.

Second, from the perspective of the social agency effect, Mayer\cite{ref:mayer-social} has shown that conversational exchanges in multimedia learning activate learning schemas and are associated with comprehension.

The design of the present system was also inspired by the Expert-$\times$-Novice structure of cognitive apprenticeship theory\cite{ref:cognitive-apprenticeship}. Collins et al.\cite{ref:cognitive-apprenticeship} systematized a series of instructional strategies---modeling, coaching, and scaffolding---and presented design principles for learning environments that promote the internalization of knowledge.

\subsection{AI-Assisted Educational Content Generation}

Research on automatically generating presentation slides from documents has evolved from rule-based and extractive summarization approaches\cite{ref:d2s} to LLM-powered approaches\cite{ref:pptagent}.
More recently, systems that generate slides and audio in an integrated manner have also been proposed\cite{ref:pass}. PPTAGENT\cite{ref:pptagent} and PASS\cite{ref:pass} combine automated structural organization of slide content with speech synthesis to streamline the generation of presentation videos.

However, these existing systems focus primarily on academic papers and emphasize technical evaluation.
Furthermore, they typically treat narration scripts as simple read-alouds or summaries of slide text, with insufficient optimization for the ``listenability'' that is specific to TTS.

\subsection{Positioning and Contributions of This Study}

Building on the above research trends, this study proposes and implements the following three contributions.

\begin{itemize}
\item \textbf{A human-in-the-loop semi-automated lesson generation system}:
  We design and implement a semi-automated generation workflow incorporating three stages of human supervision, leveraging LLMs and TTS technology with the aim of augmenting rather than replacing educators.
\item \textbf{Semi-automated generation and TTS optimization of Expert-$\times$-Novice dialogue narration}:
  We establish a novel methodology for automatically generating dialogue-format narration inspired by cognitive apprenticeship theory, and present design principles for TTS scripts optimized for listenability.
\item \textbf{Exploratory evaluation of learner-reported outcomes in actual classroom settings}:
  Through a quasi-experiment with 245 first-year high school students, we \rev{compare three lesson formats (instructor voice, single TTS, dialogue TTS) on outcomes derived from} the ARCS model, Cognitive Load Theory, and learning engagement theory\rev{, and report how the dialogue TTS session was rated differently, including the respects in which it was rated lower}.
\end{itemize}

Existing dialogue-learning research presupposes human teachers, and existing automated content generation systems (PPTAGENT, PASS, etc.) remain at the level of technical evaluation without validating educational effects on actual learners.
\rev{Recent work has begun to combine cognitive apprenticeship with generative AI from the other direction. Adler and Shani}\cite{ref:adler2026}\rev{ position the model as a partner in a hybrid apprenticeship through which pre-service teachers develop instructional practice, and Yugopuspito et al.}\cite{ref:yugopuspito2026}\rev{ report an empirical study of student perceptions and learning outcomes under generative-AI-supported apprenticeship. In both, the generative model participates in the apprenticeship relationship. The present study differs in using the model to \emph{author} the apprenticeship dialogue in advance, so that what learners receive is a fixed artifact rather than an interaction, which is what makes it deliverable at the scale of a school curriculum and also what limits it.}

\rev{The contribution is accordingly integrative and empirical rather than methodological, and we state it in those terms: the components are existing ones, and what is new is a pipeline in which narration is structured by an instructional theory, a documented account of the educator review step, and an evaluation with 245 learners in an actual curriculum.}

\section{Proposed System}

\subsection{Design Philosophy}

This system was intentionally designed not as a ``one-shot'' fully automated generation tool but as an interactive and augmentative workflow.
This means that educators are positioned as using AI as an efficient assistant while retaining final judgment over quality.

\subsection{System Architecture}

\begin{figure}[htbp]
  \centering
  \includegraphics[width=\textwidth]{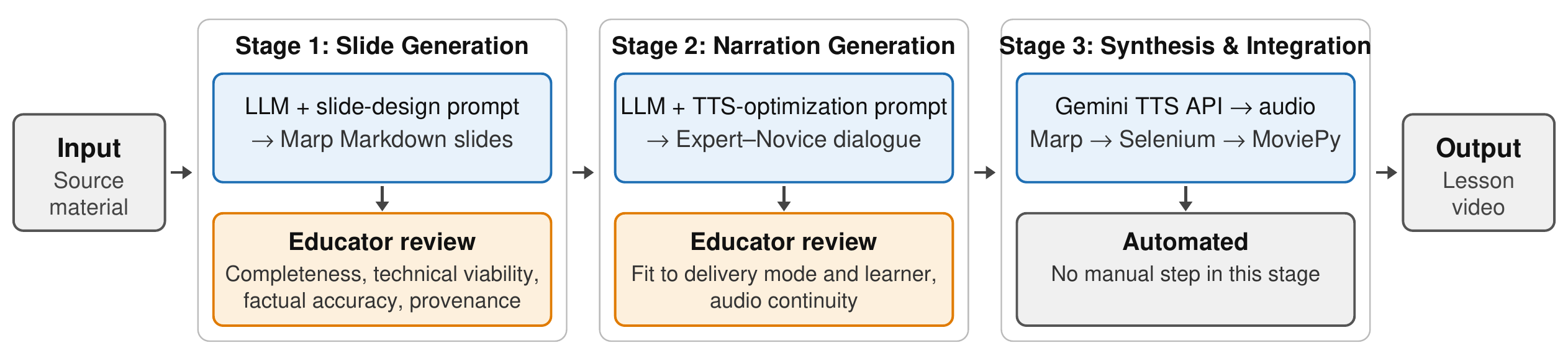}
  \caption{Architecture of the proposed system\label{fig:architecture}}
\end{figure}

As shown in Figure~\ref{fig:architecture}, the system comprised the following three main stages.

\subsubsection{Stage 1: Structured Slide Generation}

The technology stack comprised Claude 3.5 Sonnet (Anthropic, LLM) and Marp (Markdown-based slide generation).
Specifically, unstructured text such as textbook chapters or lecture materials was provided as input, and the LLM was instructed to generate slides in Marp format. The prompt included the following elements (the items listed below were composed as a single prompt):

\begin{itemize}
\item Basic configuration: initial Marp theme settings and YAML description
\item Structural rules: page breaks, use of headers, and heading level constraints
\item Design elements: markers, callout boxes, two-column layout, equations, image placement, and code blocks
\item Color palette: six theme-defined colors (background, main, accent, etc.)
\item Best practices: one message per slide, visual creativity, and overall flow
\item Implementation examples: concrete slide code
\item Notes: utilization of CSS classes, mandatory specification of figure and table sizes, etc.
\end{itemize}

As Human Feedback 1, the educator reviewed the generated slides and performed fact-checking, refinement of the pedagogical flow, and addition of figures and tables.

\subsubsection{Stage 2: TTS-Optimized Narration Generation}
A narration script was generated for each slide. Narration scripts written directly by humans tend to use written-language expressions and unnatural sentence structures, causing TTS engines to generate awkward speech. To address this problem, the system used generative AI to automatically produce scripts optimized for the ``listenability'' specific to TTS.

As common design principles, nine optimization strategies for TTS narration scripts are presented in Table~\ref{tab:tts-design-principles}.

\begin{table*}[htbp]
\caption{Design Principles for TTS Narration Scripts}
\label{tab:tts-design-principles}
\centering
\small
\begin{adjustbox}{max width=\textwidth}
\begin{tabular}{lp{10cm}}
\toprule
Item & Description \\
\midrule
Audience specification & Specify grade level and learner profile (e.g., first-year high school students, engineering undergraduates) \\
Difficulty calibration & Select vocabulary and examples appropriate to the age and profile of the target audience \\
Sentence length control & Aim for approximately 30--40 characters per sentence to maintain a length that TTS can read naturally \\
Pacing & Effective punctuation placement to create natural pauses that support viewer comprehension \\
Audio quality considerations & Disambiguation of homophones, furigana for difficult kanji, and optimization of numeral and symbol notation \\
Objective expression & Avoid personal experience and subjective expression; unify content to a factual basis \\
Linguistic expression of emotion & Express emotion and emphasis through linguistic means that remain intelligible via TTS \\
Reading rhythm & Balance short and long sentences to create a TTS-appropriate rhythm \\
Visual coordination & Coordinate with slides using expressions such as ``Please look at this figure'' or ``The red area on screen'' \\
\bottomrule
\end{tabular}
\end{adjustbox}
\end{table*}

This study provided two modes selectable according to the nature of the learning content and the learning objective.

(1) Single-speaker mode:
This mode is suited to systematic and structured one-way information delivery. It explicitly marks important points using linguistic emphasis expressions (e.g., ``What is important is\ldots'' and ``Notably\ldots''), following a structured composition of opening $\to$ introduction $\to$ main content $\to$ reinforcement $\to$ confirmation $\to$ summary $\to$ closing. Objective supplementary information such as research findings, historical facts, and application examples was inserted as appropriate to support learner comprehension.

(2) Dialogue mode (the primary innovation of this study):
Inspired by cognitive apprenticeship theory\cite{ref:cognitive-apprenticeship}, we automatically generated dialogue-format narration featuring two personas, ``Expert'' and ``Novice.'' This realizes a natural conversational flow and gives voice to learners' thought processes. The Expert is responsible for providing accurate knowledge, offering explanations through metaphors and concrete examples, and presenting advanced perspectives, while the Novice acts as a proxy for the learner by asking naive questions (``Why?'', ``What does that mean?''), rephrasing in their own words (``So you mean\ldots''), and confirming and organizing understanding. The dialogue was structured around a five-stage basic pattern---Expert introduces a concept $\to$ Novice asks a question $\to$ Expert elaborates $\to$ Novice rephrases $\to$ Expert confirms and extends---with a speech ratio of Expert 7: Novice 3, and approximately 2--3 sentences per utterance as a baseline. Five types of question strategies were used selectively: Why-type (cause), How-type (method), What-if-type (hypothetical), association, and confirmation, with scaffolding that expresses the process by which the Novice gradually deepens understanding.

This dialogue design preemptively resolves learners' latent questions and provides vicarious experience\cite{ref:bandura1977}, while also inducing self-explanation effects and realizing a collaborative knowledge-construction process grounded in social constructivism.

The correspondence between the Expert--Novice dialogue in this system and each element of cognitive apprenticeship is shown in Table~\ref{tab:apprenticeship}.

\begin{table}[htbp]
\caption{Elements of Cognitive Apprenticeship and Their Realization in Dialogue Format}
\label{tab:apprenticeship}
\centering
\small
\begin{adjustbox}{max width=\textwidth}
\begin{tabular}{p{2cm}p{5cm}}
\toprule
Element of Cognitive Apprenticeship & Realization in Dialogue Format \\
\midrule
Modeling & Verbalization of Expert's thought process \\
Coaching & Expert's responses and elaborations to Novice \\
Scaffolding & Stepwise comprehension support via 5-stage dialogue pattern \\
\bottomrule
\end{tabular}
\end{adjustbox}
\end{table}

It should be noted that this system does not faithfully implement all elements of cognitive apprenticeship; in particular, fading (gradual reduction of support) is not realized. The system borrows only the inspiration of the Expert-$\times$-Novice dialogue structure, and to make this clear, the phrase ``inspired by'' is used throughout the text.

As output, the TTS-optimized narration script for each slide was embedded in the Marp file in HTML comment format.

As Human Feedback 2, the educator reviewed the narration scripts and verified and corrected accuracy, clarity, and alignment with instructional style.

\subsubsection{Stage 3: Speech Synthesis and Integration}

In this stage, audio narration was added to the Markdown slides generated in Stage 2 and integrated as final video content. Using the audio generation capabilities of the Gemini TTS API (model: \texttt{gemini-2.5-pro-preview-tts}, google-genai SDK), we implemented two narration modes: a single-speaker lecture format and an Expert--Novice dialogue format. In each mode, prompt engineering was used to control speaker characteristics (clear teacher-like speech, a curious young person's voice, etc.) to generate audio suited to instructional listening. The generated audio was synchronized with the slide images rendered by Marp and output as a final video using MoviePy.

The following procedure was used to generate the video with audio.

(1) A Python script extracted the narration script for each slide.

(2) Audio generation requests were sent to the Gemini TTS API.
  \begin{itemize}
  \item Single-speaker mode: The prebuilt voice Enceladus was used, with an English prompt instructing clear teacher-like speech.
  \item Dialogue mode: Enceladus was assigned to Expert and Autonoe to Novice, with English prompts instructing each speaker's characteristics (expert-like speech for Expert; an inquisitive young person's voice for Novice).
  \item Generation settings: Only \rev{\texttt{response\_\allowbreak modalities}} \rev{\mbox{$=$}} \rev{\texttt{["AUDIO"]}} was specified; parameters such as sampling temperature were not set.
  \item Audio format: 24~kHz, mono, 16-bit PCM WAV; 1 second of silence was inserted before and after each segment (pydub).
  \item Prosody control: Inline tags (\texttt{[short pause]}, \texttt{[medium pause]}, \texttt{[long pause]}, \texttt{[uhm]}, etc.) were used to control pauses and intonation.
  \item The full TTS prompts are published as online supplementary material (described below).
  \end{itemize}

(3) Marp slides were converted to HTML using VSCode.

(4) Images were rendered from HTML using Selenium (SVG animations and similar embedded elements were rendered as video).

(5) Images and audio were synchronized using MoviePy to generate the final video.

The implementation code and narration script generation prompt templates for this system are publicly available at \url{https://github.com/moiku/gemini-tts-lecture-generation}.

\rev{The dialogue TTS lesson evaluated in Experiment 2 is submitted as supplementary material in full (22~min 14~s). Because the central claim of this paper concerns how synthesized narration is experienced, and because the audio characteristics we report cannot be conveyed in print, we provide the lesson as delivered rather than excerpts. Readers can hear the drift in voice characteristics between separately generated segments that is discussed in Section}~\ref{sec:audio}\rev{, and judge the Expert--Novice exchange for themselves. The video contains slides and synthesized narration only; no participant appears in it. It is in Japanese, the language of instruction, and its slides are those reproduced in Figure}~\ref{fig:lesson-slides}\rev{.}

\subsection{\rev{Educator Review: What Is Checked and What Is Changed}}
\label{sec:hitl}

\rev{The workflow places an educator between generation and delivery, but describing that step as ``quality control'' says nothing about what is actually examined or how often the output needs changing. Because both lessons used in the validation experiment were produced through this workflow, and the generation sessions, the raw model output, and the version delivered to learners were all retained, the review step can be described from the record rather than from intention. This section reports what that record shows. It concerns the workflow rather than learner outcomes, and is the basis for RQ3.}

\rev{Across the two lessons the educator issued twelve revision instructions over five rounds, and made further changes directly in the files. Table}~\ref{tab:review-criteria}\rev{ organizes these into seven criteria. The criteria were derived from the record rather than specified in advance, and we present them as a description of one team's practice, not as a validated instrument.}

\begin{table}[htbp]
\caption{Review Criteria Derived from the Generation Records}
\label{tab:review-criteria}
\centering
\small
\begin{adjustbox}{max width=\textwidth}
\begin{tabular}{@{}p{2.2cm}p{5.4cm}@{}}
\toprule
Criterion & What it checks, with an instance from the record \\
\midrule
Completeness & Whether the output is whole. One generation stopped mid-sentence partway through the second of five planned chapters and was regenerated. \\
\addlinespace
Technical viability & Whether the output renders in the delivery environment. Diagrams emitted as Mermaid markup did not render in the slide toolchain and were reissued as SVG files. \\
\addlinespace
Factual accuracy and internal consistency & Whether the material contradicts itself or overstates what is known. See the worked example below. \\
\addlinespace
Provenance & Whether claims that need a source have one. The educator supplied the official definition of the policy term used in the lesson together with its government source, which the model had not provided, and required the narration to state explicitly that the framework being taught was developed for this course rather than an established one. \\
\addlinespace
Fit to delivery mode & Whether the material suits how it will be delivered. Slides carrying timings for a live class were removed for the on-demand video, and narration directing learners to discuss in pairs was deleted because the video cannot support it. \\
\addlinespace
Fit to the learner & Whether vocabulary and explanation match the audience. English forms were added at first mention for personal and technical names, and the treatment of cognitive bias was expanded with everyday examples. \\
\addlinespace
Audio continuity & Whether the narration holds together as continuous speech. Openings that reintroduced a topic already announced on the previous slide were cut, and discourse markers that implied the wrong relation between adjacent slides were corrected. \\
\bottomrule
\end{tabular}
\end{adjustbox}
\end{table}

\rev{The last of these criteria has no counterpart in text-based generation. A slide deck is read one slide at a time, so a redundant opening sentence costs the reader little; narration is heard as an unbroken stream, and the same sentence becomes a repetition the listener cannot skip. Three of the educator's edits to the final narration were of this kind and affected no slide content.}

\begin{figure*}[htbp]
\centering
\newcommand{\slidepanel}[4]{%
  \begin{minipage}[t]{0.47\textwidth}
    \centering
    #2\\[3pt]
    {\footnotesize (#1)~#3\par}
    #4
  \end{minipage}}
\slidepanel{a}{\includegraphics[width=\linewidth]{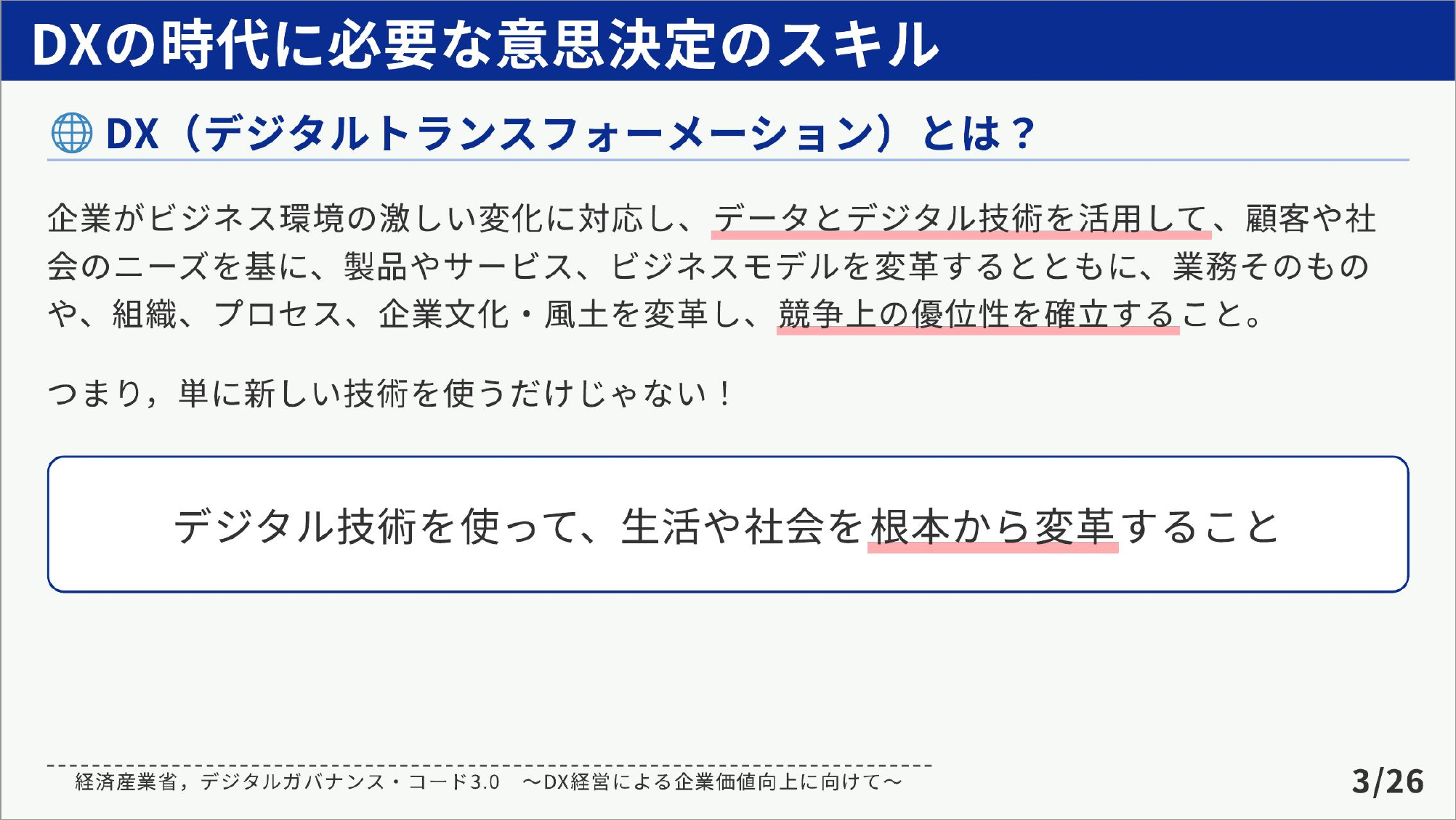}}{Definition of the policy term, with the government source the educator supplied.}{}
\hfill
\slidepanel{b}{\includegraphics[width=\linewidth]{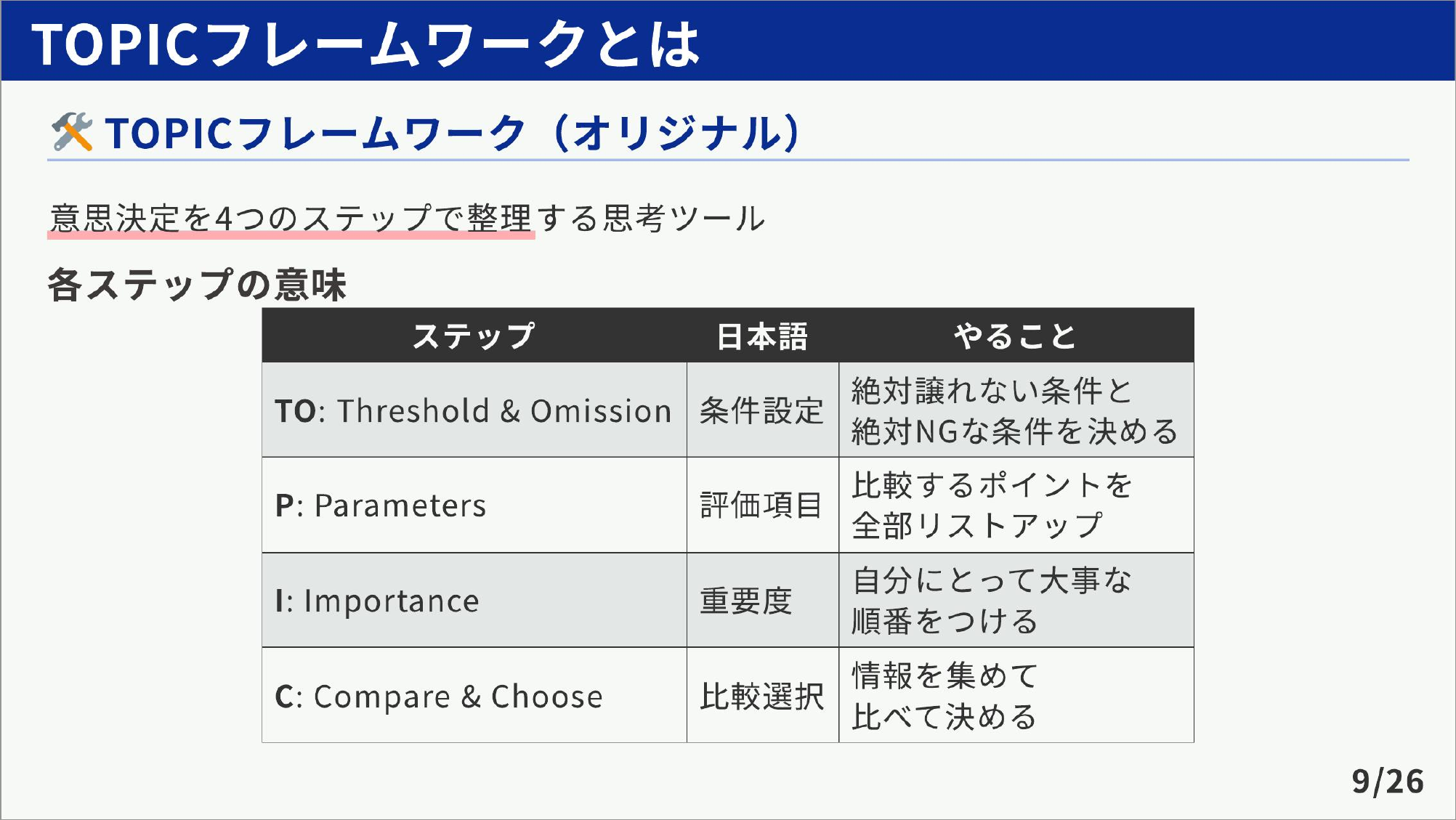}}{The framework taught, labeled \emph{original} to state that it was devised for this course.}{}

\vspace{10pt}

\slidepanel{c}{\includegraphics[width=\linewidth]{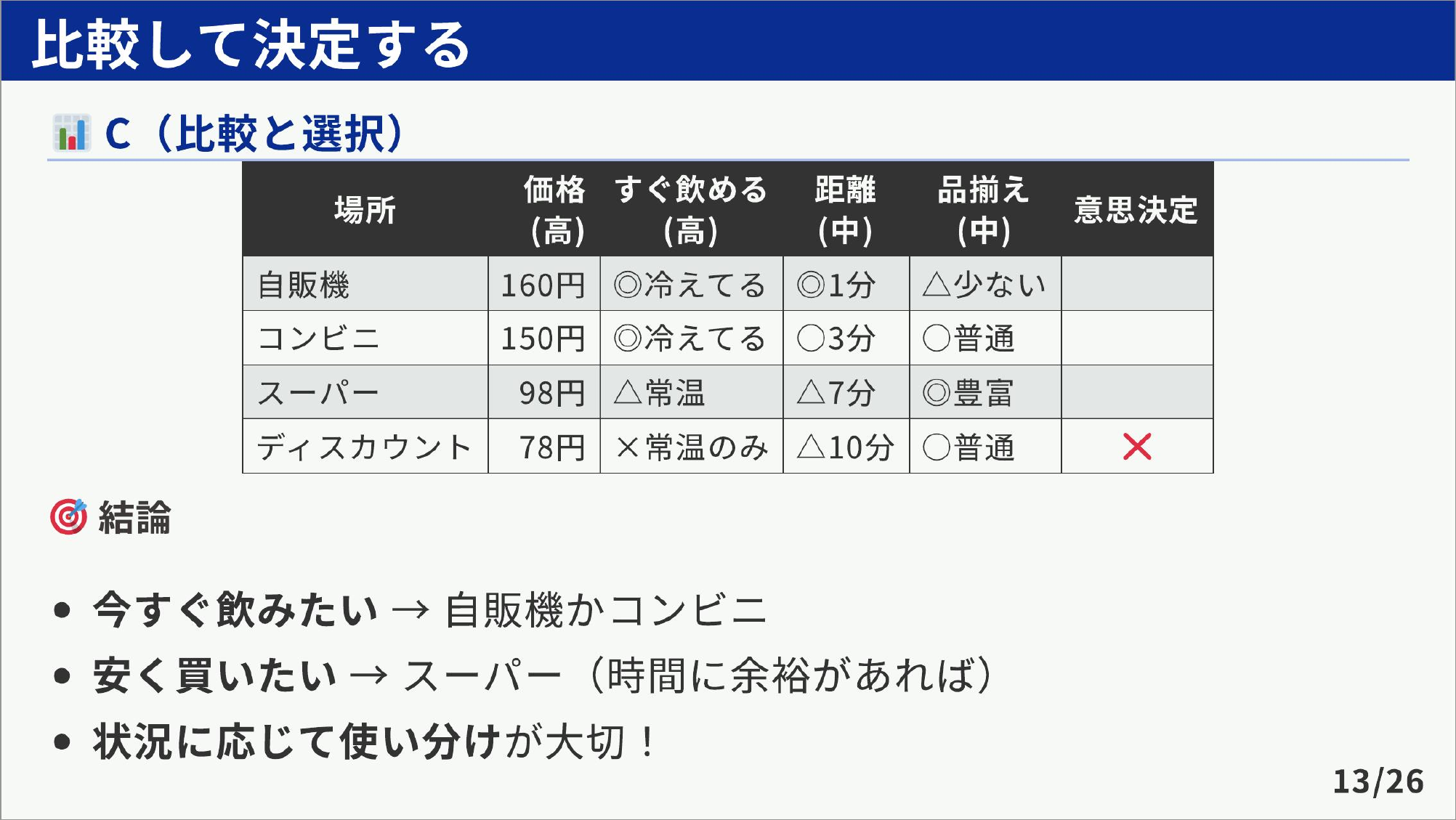}}{The comparison table after correction. Columns, left to right: place, price, chilled, distance, stock, decision. The discount store, bottom row, is recorded at ten minutes' walk---inside the threshold condition---and is excluded on the criterion it genuinely fails.}{}
\hfill
\slidepanel{d}{\includegraphics[width=\linewidth]{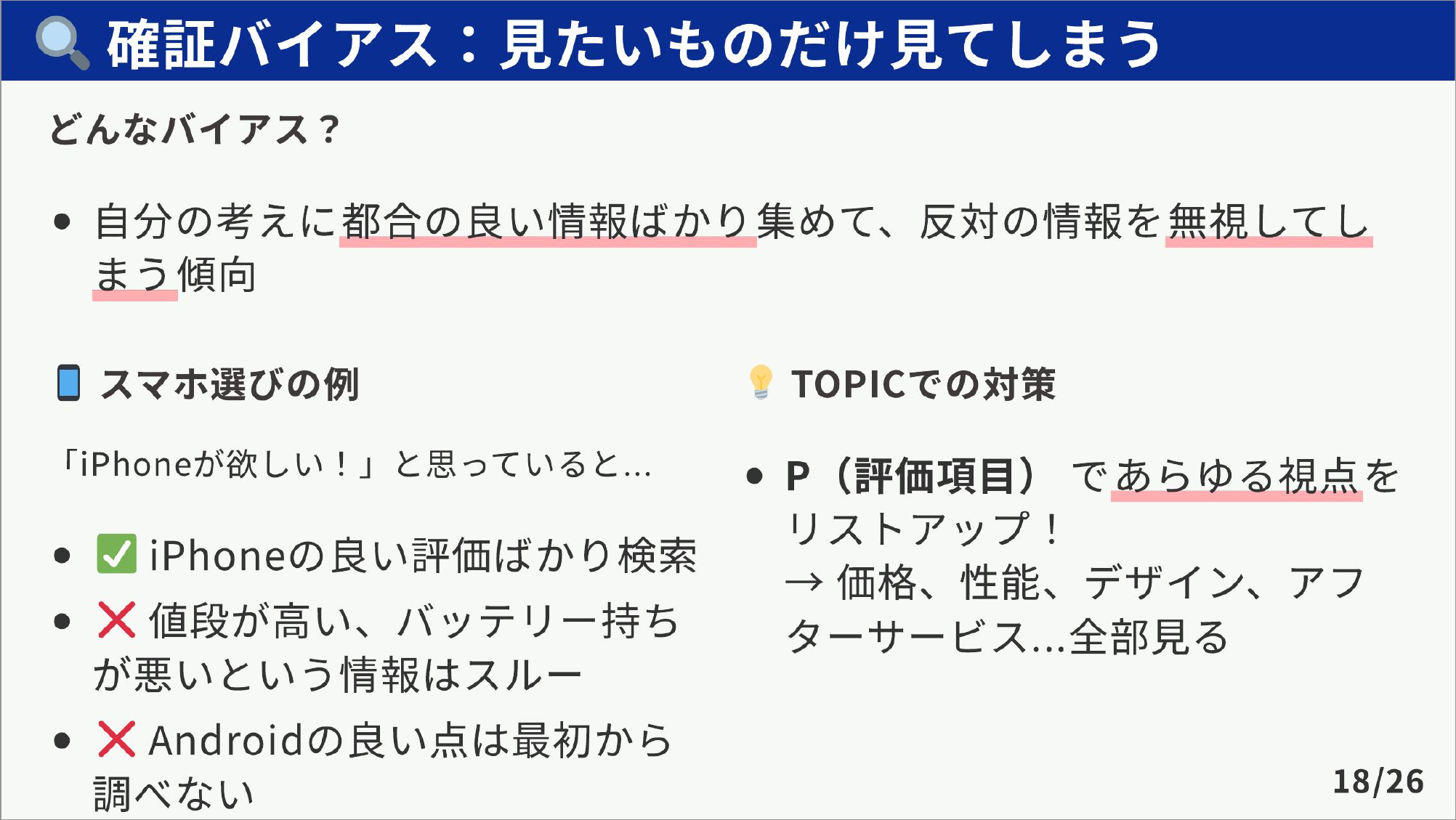}}{Cognitive bias, expanded with everyday examples on the educator's instruction.}{}

\caption{Slides from the dialogue TTS lesson as delivered, in Japanese, the language of instruction. English glosses are given beneath each panel.}
\label{fig:lesson-slides}
\end{figure*}

\rev{Figure}~\ref{fig:lesson-slides}\rev{ shows four slides from the dialogue TTS lesson as delivered, chosen to illustrate what the workflow produces and where the review step left its mark. The slides are reproduced in Japanese rather than translated, since the figure documents the material learners actually received; an English gloss accompanies each panel.}

\subsubsection{\rev{An instance of substantive correction}}
\rev{The clearest evidence that the review step does more than adjust presentation is an error it caught in the worked example of the framework being taught. The model generated an example applying the framework to choosing where to buy a drink. It set a threshold condition of being within a ten-minute walk and stated that all four candidate shops satisfied it, then presented a comparison table recording one of them as fifteen minutes away. The example thus declared a candidate admissible and simultaneously recorded it as violating the admissibility condition, and the accompanying narration excluded that candidate by appeal to the very condition it had said was met.}

\rev{The educator corrected the distance in the table and changed the stated ground of exclusion, so that the exclusion follows from a criterion the candidate genuinely fails. A related edit tightened a boundary statement elsewhere in the narration, where the model had described the threshold as excluding shops ``ten minutes or more'' away when the stated condition admits exactly ten minutes.}

\rev{Had the generated material been delivered unchanged, the lesson introducing the framework would have demonstrated it on an example that misapplied it. The error is not one of style or fluency, and it is not the kind that a reader checking for plausibility would notice: each statement is locally reasonable, and only their conjunction is inconsistent. This is the class of error that a human-in-the-loop step is supposed to catch, and in this instance it did.}

\subsubsection{\rev{Limits of this evidence}}
\rev{Two limits should be stated. The record covers two lessons produced by the authors, who are also the researchers, so it describes a practice rather than establishing how much review such material needs in general, and it cannot indicate what an educator without the same familiarity would catch. Nor can it show what was missed: the material was not independently audited, so the criteria describe what review did address, not what it should have. A study that measured review effort and residual error rate against an independent audit would be required to make the stronger claim.}

\section{Validation Experiment}

\subsection{Research Objectives and Research Questions}
\label{sec:rq}

The purpose of this experiment is to examine the utility of the proposed system and to obtain exploratory evidence regarding the acceptability of TTS audio to learners and the design of TTS lesson formats.
Based on three theoretical frameworks, this study sets the following \rev{three} research questions.

\begin{description}
  \item[RQ1] Does TTS audio substantially degrade the learning experience compared with instructor voice?
  \item[RQ2] \revb{How do learners' ratings of a dialogue-based TTS lesson (an exchange between an expert and a novice) differ from their ratings of a single-speaker TTS lesson?}
  \item[RQ3] \rev{What do educators actually review and revise in LLM-generated lesson materials, and what review criteria does that practice reveal?}
\end{description}

For RQ1, we verify whether TTS formats substantially underperform through within-subject analysis of \rev{four} core metrics across the three video formats (instructor-voice video, single TTS, and dialogue TTS).
For RQ2, we examine the differences that dialogue TTS shows relative to single TTS---and their limitations---through a multi-faceted, exploratory repeated cross-sectional comparison based on the ARCS model, Cognitive Load Theory, and learning engagement theory.
\rev{RQ2 is deliberately stated as an open question about how ratings differ rather than as a directional hypothesis that the dialogue format improves comprehension. A format that alters the delivery of a lesson may plausibly improve some aspects of the learning experience while degrading others, and an evaluation that asks only whether it improves comprehension cannot represent such a trade-off.} \revb{Because lesson format, content, order, and session length vary together between the two sessions, a difference in ratings describes the sessions as delivered and is not attributed to narration format.} \rev{We therefore report the full set of primary outcomes, including those on which the dialogue session was rated lower.}
\rev{For RQ3, we analyze the complete generation logs of the two lessons produced by the proposed system, comparing the raw LLM output, the version actually delivered to learners, and the instructions the educator issued in between. RQ3 concerns the workflow itself rather than learner outcomes, and is addressed in Section}~\ref{sec:hitl}\rev{.}
Note that the \rev{first two} RQs address different questions. RQ1 is a threshold question---``does TTS audio substantially degrade the learning experience compared with instructor voice''---while RQ2 is a \rev{descriptive question about which aspects of the learning experience the dialogue format alters, and in which direction.} Therefore, even if the two analyses yield different results, they should be interpreted as complementary rather than contradictory findings.

\subsection{Experimental Design}
\label{sec:design}

To validate the utility of the proposed system, we adopted a within-subject design in which the same group of learners ($N=245$) sequentially experienced three different lesson formats.
Participants were 245 first-year students at a prefectural high school in Niigata Prefecture, and the experiment was conducted as part of an inquiry-based learning curriculum (a data-driven career education program using generative AI) designed and delivered by the authors.

As a constraint of this experimental design, the order of implementation was fixed and the lesson content differed across sessions; therefore, the effect of lesson format cannot be completely separated from order effects and content effects. Accordingly, this study does not claim causal effects of lesson format; rather, it aims to describe the patterns observed across formats and to provide design implications derived from those patterns.

\begin{enumerate}
\item May 21, 2025: Traditional video lesson with instructor (human) voice (control condition)\\
Content: Career paths from high school graduation through university to professional life; overview of a generative-AI-powered career path proposal system.
\item June 25, 2025: Lesson generated by the proposed system using single-speaker TTS (Experiment 1)\\
Content: How generative AI works
\item July 23, 2025: Dialogue TTS lesson generated by the proposed system (Experiment 2)\\
Content: A framework for data-driven decision-making
\end{enumerate}

\rev{The running times of the three videos were 20~min 17~s (instructor voice), 41~min 47~s (single TTS), and 22~min 14~s (dialogue TTS). The single TTS session was therefore approximately twice as long as the other two. Video length is thus a further respect in which the three sessions differ, alongside content and order, and it is not separable from lesson format in this design. Session length was determined by the curriculum slot available for each lesson rather than by the format under study.}

\rev{All three lessons above were delivered as videos viewed in the students' own classroom. The curriculum also included an earlier face-to-face lesson held in the school gymnasium, and the Experiment 2 questionnaire collected retrospective ratings of that lesson as well. We excluded it from all analyses reported here. Its physical setting differed from that of the three video lessons, and the free-text responses indicate that this difference dominated the ratings: of 113 substantive comments on how the face-to-face lesson could be improved, 49 concerned heat or physical discomfort from sitting on the gymnasium floor. The influence ran in both directions, since comments praising the video lessons also cited the setting rather than the lesson (for example, 19 of 120 comments on the merits of the instructor-voice video mentioned the cooler classroom). The effect of the physical environment cannot be separated from the effect of lesson format for this condition, so including it would not have produced an interpretable comparison.}

\subsection{Theoretical Framework for Evaluation}
\label{sec:rq-framework}

To capture learner experience from multiple perspectives, this study evaluates learning from three dimensions: learning motivation, cognitive processing, and engagement. These dimensions are systematically organized by the ARCS model\cite{ref:keller-arcs}, Cognitive Load Theory\cite{ref:sweller-clt}, and learning engagement theory\cite{ref:fredricks-engagement}, respectively, and are effective for comprehensively evaluating the utility of a learning support system.

\begin{itemize}
\item ARCS model: Evaluates learning motivation from four elements---Attention, Relevance, Confidence, and Satisfaction.
\item Cognitive Load Theory: Analyzes cognitive processing from two types---extraneous load and germane load. Intrinsic load was not applied as a comparative metric between lesson formats because it depends on the complexity of the learning content.
\item Learning engagement theory: Evaluates learner engagement from two of three dimensions---emotional and cognitive---of behavioral, emotional, and cognitive engagement.
\end{itemize}

There is conceptual overlap between Attention in the ARCS model and emotional engagement, and between Confidence and cognitive engagement. In this study, these are not treated as independent constructs but rather as complementary perspectives that illuminate the same learning experience through different theoretical lenses.

\subsection{Data Collection Procedure}

A questionnaire was administered after each of Experiment 1 (single TTS) and Experiment 2 (dialogue TTS), yielding 229 (response rate 93.5\%) and 206 (response rate 84.1\%) valid responses, respectively.
Because the questionnaires were collected anonymously, individual correspondence (matching) between the two rounds is not possible.

Each questionnaire was structured as follows.

\begin{enumerate}
\item Detailed evaluation (format-specific): After each lesson, ARCS, cognitive load, and engagement items on a 5-point scale specific to that lesson format were collected ($N=229$ and $N=206$, respectively).
\item Open-ended feedback: Positive aspects and areas for improvement were collected.
\item Retrospective comparative evaluation (Experiment 2 only): At the end of the Experiment 2 questionnaire, \rev{four} metrics (ease of comprehension, \rev{sustained interest,} ease of concentration, overall evaluation) were collected in which all video formats experienced (instructor-voice video, single TTS, and dialogue TTS) were directly compared on a 5-point scale. Preference was also collected as the format perceived to have the highest learning effect (single choice), the most enjoyable format (single choice), and formats they would like to experience again (multiple choices allowed) ($N=206$; formats not experienced were excluded from evaluation).
\end{enumerate}

This secured two complementary data sources: detailed evaluation (repeated cross-sectional between-group comparison) and retrospective comparative evaluation (within-subject analysis).
The collected questionnaire items were systematically classified based on the three theoretical frameworks.
All questionnaire items were created specifically for this study. Each item was designed to operationally define the constructs of the ARCS model, Cognitive Load Theory, and learning engagement theory, and to measure them on a 5-point scale (1: strongly disagree to 5: strongly agree).
\rev{The items were written to capture situation-specific reactions to a particular lesson rather than to measure stable psychological traits, which is also the stated design intent of Keller's own ARCS-based instrument}\cite{ref:keller-arcs}\rev{. Several constructs are consequently represented by a single item. The reliability and dimensionality of the item set are examined in Section}~\ref{sec:psychometrics}\rev{, and the analyses that follow use only those composites the examination supports.}

\rev{The questionnaire also recorded two background characteristics that do not refer to the lesson content: whether respondents study using online video, and whether they had previously encountered machine-read instructional material. A third background item asked how much respondents already knew about the content of that day's lesson. Because the referent of this third item is the lesson content itself, and the content differed between the two sessions, it is treated in Section}~\ref{sec:prior-knowledge}\rev{ as a description of how far the two lessons differed rather than as a covariate.}

\section{Results and Discussion}

This section presents the results obtained from the experiment and the interpretations derived from them in an integrated manner for each theme, addressing RQ1 and RQ2 set in Section~\ref{sec:rq}.

\subsection{Analysis Framework}
\label{sec:hypotheses}

Given the constraints of the measurement design (questionnaires were collected anonymously, so the two rounds have no within-subject correspondence), we adopted a two-stage approach.
The first stage was a within-subject core analysis using the retrospective three-format comparison included in the Experiment 2 (dialogue TTS) questionnaire (direct comparison by the same respondents, $N\leq 183$), and the second stage was a repeated cross-sectional between-group comparison of the detailed items between the single TTS questionnaire ($N=229$) and the dialogue TTS questionnaire ($N=206$).
Note that although the same cohort of students likely responded to both questionnaires, individual correspondence cannot be established due to anonymous collection, so the independence of observations is not guaranteed. Because we cannot control for dependence in this study, we treat this as an approximate repeated cross-sectional comparison (see Section~\ref{sec:limitations}). \revb{The significance tests and bootstrap confidence intervals used in the second stage assume independent samples, an assumption that overlapping respondents may violate. The second stage is therefore exploratory throughout, and its results are reported as descriptions of how the two sessions were rated, not as definitive evidence of differences between narration formats.}
The correspondence between each RQ and the analytical method is as described in Section~\ref{sec:rq}.

\revb{The primary outcomes of the second stage are defined at the level of constructs rather than individual items. They were defined during the revision of this manuscript, after the item-level comparisons reported in its original version had been examined, and the composites rest on the factor analysis in Section}~\ref{sec:psychometrics}\revb{, which was conducted on the same responses. They are therefore not pre-specified outcomes, and the analytical sequence is itself a further reason to treat the second stage as exploratory.} \rev{Six composite indices were formed from the items common to both questionnaires, and six further constructs that are represented by only one common item are carried as single-item outcomes; the twelve together constitute the primary outcomes, and the Benjamini--Hochberg false discovery rate procedure is applied across them.}

\rev{Item-level comparisons are retained as a secondary, diagnostic analysis with its own correction, because not all items are interchangeable indicators of a common construct. Whether synthesized speech sounds natural, whether it is intelligible, and whether its speaking rate is appropriate are distinct properties of the generated audio, and a composite that averages them would conceal which property a given format affects. The distinction drawn here is between items that reflect a shared underlying construct, which are combined, and items that record separate design-relevant attributes, which are not.}

\rev{Effect sizes are reported as rank-biserial correlations, equivalent to Cliff's \mbox{$\delta$}, with percentile bootstrap confidence intervals from 2{,}000 resamples. The sign convention is that a positive value indicates a higher rating in the dialogue TTS session. Effect sizes are interpreted using Cohen's conventional benchmarks}\cite{ref:cohen1992}\rev{, under which \mbox{$|r| = .10$}, \mbox{$.30$}, and \mbox{$.50$} correspond to small, medium, and large effects.}

\subsection{Comparability of the Two Respondent Groups}
\label{sec:prior-knowledge}
\rev{Because the two questionnaires were collected anonymously, the comparability of the two respondent groups cannot be established by individual matching and must instead be assessed from the background items. Of these, only those whose wording does not refer to the lesson content can serve that purpose.}

\rev{Habitual use of online video for study, which does not refer to the lesson, was almost identical in the two groups (single TTS \mbox{$M=2.80$}, dialogue TTS \mbox{$M=2.82$} on a 4-point scale; \mbox{$U=23109.5$}, \mbox{$p=.694$}, \mbox{$r=+.020$}), with the four response categories occupying nearly the same proportions in each (never 6.6\% and 7.8\%, rarely 25.8\% and 22.3\%, sometimes 48.9\% and 50.0\%, often 18.8\% and 19.9\%). To the extent that this item characterizes the respondents, the two groups are drawn from the same population, as would be expected given that both questionnaires were administered to the same cohort.}

\rev{Reported prior exposure to machine-read instructional material, by contrast, rose sharply between the two rounds, from 44.5\% to 69.9\% answering that they had encountered such material (\mbox{$\chi^2(1)=27.37$}, \mbox{$p<.001$}). This shift is a direct consequence of the study itself: the single TTS lesson was for many students their first encounter with machine-read instruction, so by the time of the second questionnaire it had become part of their experience. The item is therefore not a baseline characteristic but a record of carry-over between sessions, and it is reported here as such rather than used to adjust any comparison.}

\subsubsection{Prior knowledge as a measure of how far the two lessons differed}
\label{sec:prior}
\rev{The remaining background item asked how much respondents already knew about the content of that day's lesson. It was worded identically in both questionnaires, but what it refers to is not: in the first session the content was how generative AI works, a subject saturated in popular media, whereas in the second it was a decision-making framework devised specifically for this curriculum, which no student could have encountered beforehand. The responses reflect this. In the dialogue TTS session 66.0\% reported knowing nothing at all about the content, against 34.1\% in the single TTS session (\mbox{$\chi^2(1)=43.05$}, \mbox{$p<.001$}; \mbox{$U=31369$}, \mbox{$p<.001$}, \mbox{$r=-.330$}, a medium-sized difference).}

\rev{We therefore do not treat this item as a covariate, for three reasons. First, identical wording does not make it the same measurement when the referent differs; a student reporting no prior knowledge of a framework invented for the lesson is stating something close to a tautology, and no common scale links that response to a report about generative AI. Second, lesson format and lesson content are completely confounded in this design, so prior knowledge is determined by which content was selected and is a consequence of the exposure rather than a characteristic that precedes it; conditioning on it removes part of the very variation under study. Third, the item is strongly related to group (\mbox{$r=-.330$}) but barely related to the outcomes, the median absolute Spearman correlation with the questionnaire items being \mbox{$.081$} in the single TTS group and \mbox{$.064$} in the dialogue TTS group, with only 2 of 21 and 1 of 30 items exceeding \mbox{$.20$}. A variable with that profile does not remove confounding, and adjusting for it can instead amplify whatever unmeasured confounding remains. A further practical obstacle is that 136 of 206 respondents in the dialogue TTS group fall in the lowest category, leaving little variation to adjust on.}

\rev{What the item does measure well is how different the two lessons were. That two thirds of one group and one third of the other began with no prior knowledge is a direct indication of how far the contents diverged, and it is in this descriptive sense that we report it. An analysis that does adjust for prior knowledge is reported in Section}~\ref{sec:olr}\rev{ as a sensitivity analysis, with the above reservations attached. For completeness, subjective difficulty (``the lesson content was difficult'') did not differ between the groups (\mbox{$U=24959$}, \mbox{$p=.270$}, \mbox{$r=-.058$}).}

\subsection{\rev{Reliability and Dimensionality of the Questionnaire}}
\label{sec:psychometrics}

\rev{The items were written for this study, so their measurement properties cannot be assumed and are examined here before they are used. The examination is restricted to the items common to both questionnaires, since only those can support a between-group comparison. Because the items are five-point ordinal ratings, reliability is reported both as the conventional coefficient computed on raw scores and as the ordinal coefficient computed from polychoric correlations; the latter does not attenuate the estimate by treating ordered categories as interval measurements. Confidence intervals are percentile bootstrap intervals. Indices formed from two items are reported with the inter-item correlation and the Spearman--Brown coefficient rather than as validated scales.}

\begin{table}[htbp]
\caption{Reliability of the Six Composite Indices ($N=229$ single TTS, $N=206$ dialogue TTS)}
\label{tab:reliability}
\centering
\footnotesize
\setlength{\tabcolsep}{4pt}
\begin{adjustbox}{max width=\textwidth}
\begin{tabular}{@{}lccccc@{}}
\toprule
\multirow{2}{*}{Index} & \multirow{2}{*}{Items} & \multicolumn{2}{c}{Single TTS} & \multicolumn{2}{c}{Dialogue TTS} \\
\cmidrule(lr){3-4}\cmidrule(lr){5-6}
 & & $\alpha$ & $\alpha_\mathrm{ord}$ & $\alpha$ & $\alpha_\mathrm{ord}$ \\
\midrule
Attention          & 3 & .888 & .921 & .902 & .936 \\
Relevance          & 2 & .784 & .865 & .829 & .897 \\
Satisfaction       & 2 & .857 & .915 & .932 & .983 \\
Extraneous load    & 2 & .674 & .731 & .742 & .803 \\
Perceived learning & 3 & .708 & .774 & .782 & .864 \\
Audio quality      & 3 & .703 & .765 & .767 & .834 \\
\bottomrule
\addlinespace[2pt]
\multicolumn{6}{@{}p{\columnwidth}@{}}{\footnotesize $\alpha$: Cronbach's $\alpha$ on raw scores. $\alpha_\mathrm{ord}$: ordinal $\alpha$ from polychoric correlations. For the three-item indices, ordinal McDonald's $\omega$ was within $.013$ of $\alpha_\mathrm{ord}$ throughout.} \\
\end{tabular}
\end{adjustbox}
\end{table}

\rev{Table}~\ref{tab:reliability}\rev{ reports the results. Every composite reaches the conventional threshold of }$.70$\rev{ in both sessions on the ordinal coefficient. Two entries deserve comment rather than reassurance. Extraneous load is the weakest index, and in the single TTS session the lower bound of its bootstrap interval falls below the threshold (95\% CI }$[.623, .817]$\rev{), so it should not be described as well established. At the other extreme, the two satisfaction items in the dialogue TTS session correlate so highly (}$\alpha_\mathrm{ord} = .983$\rev{) that the value is better read as redundancy between near-duplicate items than as evidence of a well-constructed scale.}

\rev{Six further constructs---confidence, intrinsic load, germane load, and the behavioral, emotional, and cognitive facets of engagement---are represented by a single common item each. No reliability coefficient is computed for them, and they are carried through the analysis as single items whose measurement error is unknown. This is a limitation of the instrument, not a property of the constructs.}

\subsubsection{\rev{Factor structure}}
\rev{Sampling adequacy was acceptable in both sessions (KMO }$= .878$\rev{ and }$.909$\rev{; Bartlett's test of sphericity }$p < .001$\rev{ in both). Because the data are ordinal, the analysis used polychoric correlations, and the number of factors was determined by parallel analysis against random data generated to reproduce each item's observed category proportions. Extraction used the minimum residual method with promax rotation. Three factors were retained in both sessions.}

\rev{The same three-dimensional structure emerged in each. The first and largest factor gathers interest, relevance, satisfaction, confidence, enjoyment, and perceived learning, accounting for about 30\% of the variance in both sessions; it is best read as a general evaluation of the learning experience. The second collects the audio quality items together with the reverse-worded item on audio-related difficulty, and in the dialogue session it also collects the items concerning the Novice persona. The third comprises perceived difficulty.}

\rev{This structure does not correspond to the divisions of the ARCS model, Cognitive Load Theory, and engagement theory from which the items were drawn: the four ARCS subscales do not separate, and relevance and satisfaction in particular load together. We report this as a property of the present item set rather than of the underlying theories, for two reasons. The instrument here is a shortened and modified one in which confidence is measured by a single item, and the analysis pools ARCS items with cognitive load, audio quality, and perceived learning items, so it is not a test of the four-factor ARCS structure. Keller also cautions against expecting the intended factor structure to emerge from instruments of this kind, since they are designed to capture situation-specific attitudes rather than psychological constructs}\cite{ref:keller-arcs}\rev{. What can be said is that the discriminant validity of the four subscales was not established for these items, and the analyses that follow accordingly avoid claims at the level of individual ARCS components.}

\rev{Two further caveats attach to the factor analysis. The polychoric correlation matrix for the dialogue TTS session was not positive definite (smallest eigenvalue }$-.016$\rev{) and was smoothed to the nearest positive definite matrix, with a maximum change of }$.011$\rev{ in any element. The ratio of respondents to items in that session is }$206:30$\rev{, which is adequate for a three-factor solution but not generous. The factor analysis is therefore reported as exploratory support for the composites used below, not as a validated measurement model.}

\subsection{RQ1: Does TTS Audio Substantially Degrade the Learning Experience?}
\label{sec:friedman}

In the comparative evaluation portion of the Experiment 2 (dialogue TTS) questionnaire, the same respondents retrospectively rated all video formats they had experienced on a 5-point scale.
The results of the Friedman test on core evaluation metrics for the three video formats are shown in Table~\ref{tab:friedman-results}.
\rev{All three lessons compared here were viewed in the students' own classroom, so the delivery setting is held constant and the formats differ in the voice that narrates them.}

No significant differences between lesson formats were found on any of the core metrics.
\rev{Kendall's coefficient of concordance, which expresses how much of the ranking variance is attributable to the format, did not exceed }$.011$\rev{ on any metric, indicating that format accounted for almost none of it.}
Furthermore, to confirm that non-significance was not a false negative due to insufficient statistical power, TOST (Two One-Sided Tests) equivalence testing was supplementally conducted with equivalence bounds $\Delta=\pm0.5$ points. \rev{Each TTS format was compared with the instructor-voice video on each of the four metrics, and all eight comparisons fell within the equivalence range (all }$p<.0001$\rev{). The widest of the eight 90\% confidence intervals for the mean difference was }$[-0.212, +0.015]$\rev{, well inside the equivalence bounds; the largest absolute mean difference was }$0.098$\rev{ points on a five-point scale.}

\begin{table*}[htbp]
\caption{Within-Subject Comparison of Four Core Metrics (Friedman Test)}
\label{tab:friedman-results}
\centering
\small
\begin{adjustbox}{max width=\textwidth}
\begin{tabular}{lccccccc}
\toprule
\multirow{2}{*}{Metric} & \multicolumn{3}{c}{Mean} &
\multirow{2}{*}{$N$} & \multirow{2}{*}{$\chi^2(2)$} &
\multirow{2}{*}{$p$} & \multirow{2}{*}{$W$} \\
\cmidrule(lr){2-4}
 & Instructor video & Single TTS & Dialogue TTS & & & & \\
\midrule
Ease of comprehension & 3.55 & 3.51 & 3.51 & 182 & 0.28 & .869 & .001 \\
Sustained interest    & 3.32 & 3.25 & 3.33 & 183 & 1.84 & .398 & .005 \\
Ease of concentration & 3.29 & 3.23 & 3.25 & 183 & 1.53 & .466 & .004 \\
Overall evaluation    & 3.42 & 3.32 & 3.35 & 183 & 3.91 & .142 & .011 \\
\bottomrule
\multicolumn{8}{l}{\footnotesize $W$: Kendall's coefficient of concordance (effect size).} \\
\end{tabular}
\end{adjustbox}
\end{table*}

Concerns that replacing educational video audio with TTS would degrade the learner experience represent the greatest barrier to TTS adoption.
The within-subject analysis above found no significant differences \rev{on any of the four metrics} even when the same learners retrospectively compared all three formats.
This result suggests that TTS audio does not undermine the fundamental learning experience of a lesson, supporting the legitimacy of TTS use in educational content production.

\rev{The free-text responses suggest why the ratings converge. Asked what was good about the machine-read lesson, students described properties that follow from synthesis itself: that being machine speech it is never rendered unintelligible by poor articulation, that there is no stumbling over words, that the volume is constant, and that it makes no mistakes. Asked the same about the instructor-voice lesson, they described the opposite kind of property: that a human can improvise, bring in an analogy on the spot, and respond to the moment, and that there was feeling in the delivery. The two formats appear to be rated on different strengths that offset one another in a global judgment.}

\rev{This converges with the one directly comparable experiment we are aware of. Deng et al.}\cite{ref:deng2026}\rev{ randomly assigned undergraduates to human or synthetic narration with content and duration held constant, and found no effect of voice type on motivation or on knowledge retention measured by test. Their design controls what ours cannot, and reaches the same conclusion about voice from an independent sample in a different country and language. Two studies with complementary weaknesses agreeing is a stronger basis for the acceptability claim than either alone.}

\rev{The attributions made to the instructor-voice lesson warrant a note, since that lesson was also a pre-recorded video and could not in fact respond to anyone. Responsiveness was nonetheless attributed to it. Rather than describing a property of the recording, these responses are more plausibly read as expectations evoked by the knowledge that the voice is human, which is consistent with accounts of social cues in multimedia learning}\cite{ref:mayer-social}\rev{. Whether a synthesized voice can evoke comparable expectations is not answered by the present data.}

However, two caveats are required for this interpretation.
First, in the context of inquiry-based learning for first-year high school students, the relevance of the day's content to oneself and the difficulty level of each unit are likely to function as primary drivers of evaluation, potentially overshadowing format differences.
Second, individual differences were a larger source of variance than differences between video formats, suggesting the need to explicitly control for individual-difference variables such as learning styles and technology experience in future research.

Note that the absence of a significant difference is not itself positive evidence of equivalence. To reinforce this point, TOST equivalence testing confirmed that \rev{all four core metrics} fell within the equivalence range at bounds $\Delta=\pm0.5$ points, supporting the conclusion that non-significance is unlikely to be a false negative due to insufficient power. However, this retrospective evaluation is based on recall at the time of Experiment 2 (after experiencing dialogue TTS), with the instructor-voice video approximately two months prior and the single TTS approximately one month prior; thus, memory smoothing due to temporal distance may have occurred. In addition, since the lesson content differed across the three formats, what this test demonstrates is not ``equivalence of audio format'' but rather the fact that ``retrospective evaluations of sessions with different content, timing, and format fell within $\pm0.5$ points.'' The equivalence bound of $\Delta=\pm0.5$ is also based on a conventional criterion of 10\% of a 5-point scale, and justification based on a minimally educationally meaningful difference remains a topic for future work.
\rev{Since recording narration is the step that makes producing lesson videos expensive, this result bears directly on whether the system is worth using; we return to it in Section}~\ref{sec:implications}\rev{. What TTS additionally makes available, at no further cost, is control over how many speakers a lesson has, and that is the question examined next.}

\subsection{RQ2: \rev{What the Dialogue Format Changes}}
\label{sec:rq2}

Unlike instructor voice, TTS can introduce multiple speakers at no additional cost.
In this study, we leveraged this characteristic to design Expert-$\times$-Novice dialogue-format narration inspired by cognitive apprenticeship theory.
\rev{The comparison between the single TTS questionnaire (}$N=229$\rev{) and the dialogue TTS questionnaire (}$N=206$\rev{) is a repeated cross-sectional one}\revb{ and, because respondents likely overlap without being matched, an exploratory one;}\rev{ it is reported at the level of the twelve primary outcomes defined in Section}~\ref{sec:hypotheses}\rev{: the six composite indices whose reliability was examined in Section}~\ref{sec:psychometrics}\rev{, and the six constructs measured by a single common item. The Benjamini--Hochberg procedure is applied across these twelve.}

\begin{table*}[htbp]
\caption{Exploratory Between-Session Comparison of the Twelve Primary Outcomes}
\label{tab:primary-outcomes}
\centering
\small
\begin{adjustbox}{max width=\textwidth}
\begin{tabular}{llccccc}
\toprule
Outcome & Items & Single TTS & Dialogue TTS & $r$ [95\% CI] & $q$ & \\
\midrule
\multicolumn{7}{l}{\textit{Composite indices}} \\
ARCS: attention          & 3 & 3.294 & 3.346 & $+.034$ $[-.075, +.138]$ & .639 & \\
ARCS: relevance          & 2 & 3.825 & 3.784 & $-.035$ $[-.141, +.071]$ & .639 & \\
ARCS: satisfaction       & 2 & 3.576 & 3.711 & $+.117$ $[+.013, +.218]$ & .062 & \\
Extraneous cognitive load & 2 & 2.319 & 2.597 & $+.171$ $[+.067, +.276]$ & .010 & $^{**}$ \\
Perceived learning       & 3 & 3.488 & 3.595 & $+.093$ $[-.012, +.200]$ & .150 & \\
Audio quality            & 3 & 3.642 & 3.405 & $-.182$ $[-.288, -.079]$ & .010 & $^{**}$ \\
\midrule
\multicolumn{7}{l}{\textit{Single-item constructs}} \\
ARCS: confidence         & 1 & 3.633 & 3.820 & $+.130$ $[+.040, +.215]$ & .025 & $^{*}$ \\
Intrinsic cognitive load & 1 & 2.886 & 2.786 & $-.058$ $[-.161, +.045]$ & .405 & \\
Germane cognitive load   & 1 & 3.197 & 3.228 & $+.012$ $[-.091, +.117]$ & .892 & \\
Behavioral engagement    & 1 & 4.022 & 4.029 & $-.005$ $[-.099, +.090]$ & .918 & \\
Emotional engagement     & 1 & 3.183 & 3.383 & $+.110$ $[+.005, +.211]$ & .073 & \\
Cognitive engagement     & 1 & 3.428 & 3.626 & $+.121$ $[+.023, +.221]$ & .057 & \\
\bottomrule
\multicolumn{7}{l}{\footnotesize $^{*}q<.05$, $^{**}q<.01$ (BH-FDR corrected across the twelve outcomes). Values under each format are means.} \\
\multicolumn{7}{l}{\footnotesize $r$: rank-biserial correlation (Cliff's $\delta$); positive values indicate a higher rating in the dialogue TTS session.} \\
\multicolumn{7}{l}{\footnotesize Confidence intervals are percentile bootstrap intervals from 2{,}000 resamples. $N_\mathrm{single}=229$, $N_\mathrm{dialogue}=206$.} \\
\multicolumn{7}{l}{\footnotesize Tests and intervals assume independent samples; respondents likely overlap between sessions, so they are descriptive, not confirmatory.} \\
\end{tabular}
\end{adjustbox}
\end{table*}

\rev{Table}~\ref{tab:primary-outcomes}\rev{ reports the results. Three of the twelve outcomes differ after correction, and they do not point in a single direction. The dialogue TTS session was rated higher on confidence, that is, on believing one had understood the main content (}$r=+.130$\rev{). It was rated lower on audio quality (}$r=-.182$\rev{) and higher on extraneous cognitive load (}$r=+.171$\rev{), both to the dialogue session's disadvantage. All three effects are small by Cohen's benchmarks, and the confidence intervals for all three exclude zero by a modest margin.} \revb{Since those intervals assume independent samples, the margin should not be read as firm evidence of a difference.}

\rev{The remaining nine outcomes do not differ. This includes perceived learning (}$q=.150$\rev{), the composite that most directly represents whether learners felt they had learned, and attention, relevance, and the behavioral, emotional, and cognitive facets of engagement. Satisfaction (}$q=.062$\rev{), cognitive engagement (}$q=.057$\rev{), and emotional engagement (}$q=.073$\rev{) fall just short of the threshold. Their bootstrap intervals, which are not adjusted for multiplicity, do exclude zero by a narrow margin; the adjusted }$q$\rev{ values are the criterion we apply, and we do not interpret these three as differences.}

\rev{Read together, these results do not support a general advantage for the dialogue format.} \revb{What they describe is a mixed pattern: the dialogue session was associated with greater confidence in one's own understanding and with poorer ratings of the audio. Because the sessions also differed in content, order, and length, the pattern is not attributed to narration format. Whether a format showing such a pattern is worth adopting depends on what a lesson is for, and we return to this in Section}~\ref{sec:implications}\rev{.}

\subsubsection{\rev{Item-level diagnostic analysis}}
\label{sec:confidence}

\rev{The composite indices establish whether a construct differs; they do not identify which property of the lesson produced the difference. Table}~\ref{tab:item-level}\rev{ therefore reports the item-level comparisons, corrected separately within this block, for the seven of twenty-one items that differ.}

\begin{table*}[htbp]
\caption{Item-Level Diagnostic Comparison (Items Differing After Correction)}
\label{tab:item-level}
\centering
\small
\begin{adjustbox}{max width=\textwidth}
\begin{tabular}{llccccc}
\toprule
Item & Construct & Single TTS & Dialogue TTS & $r$ & $q$ & \\
\midrule
The synthesized voice sounded natural            & Audio quality      & 3.432 & 2.990 & $-.238$ & $<.001$ & $^{***}$ \\
The audio was hard to hear, so I could not concentrate & Extraneous load & 2.188 & 2.573 & $+.201$ & .002 & $^{**}$ \\
The synthesized voice was easy to hear           & Audio quality      & 3.852 & 3.592 & $-.155$ & .016 & $^{*}$ \\
I could explain the main content to a friend     & Perceived learning & 2.913 & 3.189 & $+.152$ & .022 & $^{*}$ \\
I think I understood the main content            & Confidence         & 3.633 & 3.820 & $+.130$ & .026 & $^{*}$ \\
What I learned will be useful in the future      & Relevance          & 4.009 & 3.825 & $-.124$ & .040 & $^{*}$ \\
The time spent on this lesson was worthwhile     & Satisfaction       & 3.498 & 3.680 & $+.126$ & .040 & $^{*}$ \\
\bottomrule
\multicolumn{7}{l}{\footnotesize $^{*}q<.05$, $^{**}q<.01$, $^{***}q<.001$ (BH-FDR corrected across the twenty-one common items).} \\
\multicolumn{7}{l}{\footnotesize The fourteen items not listed did not differ after correction.} \\
\end{tabular}
\end{adjustbox}
\end{table*}

\rev{Three points follow from the item-level results. First, the audio disadvantage is not confined to naturalness. Naturalness shows the largest effect in the entire analysis (}$r=-.238$\rev{), but ease of hearing also differs (}$r=-.155$\rev{), and so does the item asking whether difficulty hearing the audio impeded concentration (}$r=+.201$\rev{). The dialogue format was experienced as harder to listen to, not merely as less natural.}

\rev{Second, the two items that carry the confidence and perceived-learning results are about believing one has understood (}$r=+.130$\rev{) and believing one could explain the content to a friend (}$r=+.152$\rev{), whereas the item asking directly whether the main points were understood does not differ (}$q=.384$\rev{). What rose is the learner's assessment of their own grasp of the material, which is not the same thing as the grasp itself and was not measured independently in this study. The distinction matters for how the finding should be read, and we keep it explicit throughout.}

\rev{Third, one item favors the single TTS session: learners were more likely to judge that what they had learned would be useful in the future (}$r=-.124$\rev{). Since the two sessions taught different subjects, the most economical reading is that students judged how generative AI works to be more useful to them than a decision-making framework, which says more about the content than about the narration format.}

\subsubsection{\rev{The audio trade-off and where it comes from}}
\label{sec:audio}

\rev{Both the audio quality index and the extraneous load index are worse in the dialogue session, and the free-text responses indicate that neither is straightforwardly a property of the dialogue format.}

\rev{Complaints about prosody were the most common audio criticism in both sessions and at nearly the same rate, so they cannot account for the difference between them. Students in both sessions objected to unnatural intonation, and one noted a misread compound word. This is a limitation of the synthesis engine, present whichever narration mode it is used in.}

\rev{What is more frequent in the dialogue session is instability of the voice itself: students reported that the voice changed partway through and that the audio wavered. This is a known property of the model used here. Audio is generated per slide, and the preview version of the API does not guarantee identical voice characteristics across separate generation calls. With one speaker the drift is tolerated as variation within a single voice; with two, it disturbs the cues by which speakers are told apart, and resolving that ambiguity consumes resources that the lesson content would otherwise have}\cite{ref:sweller-clt}\rev{. The instructors were aware of this before delivery and had the narration disclose it to learners. It is worth noting that speaker discrimination was not itself rated as difficult (}$M=3.99$\rev{, with 5.3\% disagreeing), so the cost appears to lie in the moments where the voice shifts rather than in telling the two personas apart.}

\rev{A third source lies outside the audio altogether. The dialogue session was delivered in a forty-five-minute slot that also contained a fifteen-minute exercise with a fill-in worksheet, and complaints about slides advancing too quickly to write, and about the worksheet, were twice as frequent as in the single TTS session. One student wrote that the volume of writing left them barely able to listen to the narration at all. Dividing attention between transcription and listening would elevate the same item regardless of how the audio was produced.}

\rev{A measurement caveat compounds this. The two questionnaires word the audio-difficulty item slightly differently, the first asking whether it was hard to concentrate and the second whether concentration was impossible, so the two are not strictly the same item. The direction of the difference runs against what that wording change would predict, which is some reassurance, but the comparison is weaker than the others.}

\rev{Taken together, the elevated extraneous load in the dialogue session should not be attributed to the dialogue format as such. It is consistent with a synthesis engine whose voice characteristics drift between generation calls, compounded by the instructional design of that particular session. Both are addressable: models that reproduce consistent voice characteristics across calls, such as Qwen2.5-Omni}\cite{ref:qwen-tts}\rev{, would remove the first, and displaying speaker names or subtitles would provide a visual channel for speaker identity that does not depend on the audio at all.}

\subsubsection{Function and Limitations of the Novice Role}
\label{sec:novice}

In the dialogue design, the novice was positioned as a learner proxy, but only 39.8\% affirmed ``the novice's questions were similar to questions I had myself,'' suggesting that the novice role did not fully function as a learner proxy.
On the other hand, 57.3\% affirmed ``the novice's questions were helpful for understanding the content,'' suggesting that even if the novice's questions were limited as a vicarious experience through emotional empathy\cite{ref:bandura1977}, they functioned cognitively as scaffolding for learning\cite{ref:mayer-social}.
The low rate of question alignment may be partly attributable to the low sense of social presence of machine voice as posited by Garrison et al.'s\cite{ref:garrison2000} theory of social presence, but the accuracy of question selection and the characteristics of the content (divergence from learners' actual questions) may also have contributed. Since alignment of question content is not guaranteed even with human voice, this divergence cannot be attributed solely to machine voice.
This result suggests that the effect of dialogue format in a TTS environment relies more on cognitive scaffolding mechanisms than on social presence, implying that designing the novice role should prioritize cognitively effective questions over emotional empathy.
\rev{In the factor analysis reported in Section}~\ref{sec:psychometrics}\rev{, the items concerning the Novice persona did not form a dimension of their own but loaded with the items about how the audio and the exchange were delivered, which is consistent with their operating as a feature of the presentation rather than as a distinct channel of social engagement. Because these items were asked only in the dialogue session, they cannot be compared across formats, and we make no claim that a novice-specific dimension exists.}

\subsubsection{\rev{Sensitivity Analysis: Adjusting for Prior Knowledge}}
\label{sec:olr}

\rev{Section}~\ref{sec:prior}\rev{ set out why prior knowledge is not used as a covariate in the primary analysis. Because the previous version of this work reported such an adjustment, and because readers may reasonably wish to see it, we report it here as a sensitivity analysis, subject to those reservations.}

\rev{For the six single-item outcomes, a proportional odds model (ordinal logistic regression, \texttt{statsmodels} v0.14.6) was fitted with prior knowledge as a covariate; for the six composite indices, a rank-transformed analysis of covariance was used. Both pool the 435 responses from the two sessions and do not presuppose individual correspondence between them (unique learners: up to 245). The odds ratio of an ordinal logit expresses the tendency to respond in a higher category and is not a ratio of means.}

\rev{Under adjustment, more outcomes reach significance than in the primary analysis. Confidence strengthens (}$\mathrm{OR}=2.18$\rev{, }$q=.001$\rev{), and emotional engagement (}$\mathrm{OR}=1.62$\rev{, }$q=.019$\rev{) and cognitive engagement (}$\mathrm{OR}=1.73$\rev{, }$q=.012$\rev{) become significant, as do perceived learning (}$q=.016$\rev{) and satisfaction (}$q=.045$\rev{) among the indices. The audio quality and extraneous load results persist (}$q=.013$\rev{ and }$q=.005$\rev{).}

\rev{This pattern is the reason we do not present the adjusted analysis as the primary one. Prior knowledge is strongly associated with session but only weakly with the outcomes, and an adjustment on such a variable that moves five results from non-significant to significant is more plausibly absorbing the difference between the sessions than removing noise from it. Since the two sessions taught different subjects, part of what the adjustment removes is the content difference that is inseparable from the format difference under study. We therefore report these results as showing what follows if one is willing to treat prior knowledge as a nuisance variable, and not as evidence that the dialogue format has effects the primary analysis failed to detect.}

\subsection{Lesson Format Preference Evaluation}
\label{sec:preference}

The results of comparing learning effect, enjoyment, and desire to continue for the three video formats (instructor-voice video, single TTS, and dialogue TTS) plus a gymnasium in-person session across four formats at the end of the Experiment 2 questionnaire are shown in Table~\ref{tab:preference-results} (``highest learning effect'' and ``most enjoyable'' are proportions among the three video formats excluding the gymnasium session; $\chi^2$ is a goodness-of-fit test with a uniform distribution as the null hypothesis).

\begin{table*}[htbp]
\caption{Lesson Format Preference Results (\%)}
\label{tab:preference-results}
\centering
\small
\begin{adjustbox}{max width=\textwidth}
\begin{tabular}{lccc}
\toprule
Item & Instructor video & Single TTS & Dialogue TTS \\
\midrule
Perceived highest learning effect & 34.0 & 18.7 & 47.3 \\
(single choice, $N=150$) & \multicolumn{3}{c}{$\chi^2(2)=18.52$, $p<.001^{***}$} \\
\midrule
Most enjoyable to learn from & 15.6 & 17.5 & 66.9 \\
(single choice, $N=154$) & \multicolumn{3}{c}{$\chi^2(2)=78.09$, $p<.001^{***}$} \\
\midrule
Would like to experience again & 27.7 & 22.3 & 50.5 \\
(multiple choices, $N=206$) & \multicolumn{3}{c}{(reference only, no significance test)} \\
\bottomrule
\multicolumn{4}{l}{\small $^{***}p<.001$}
\end{tabular}
\end{adjustbox}
\end{table*}

Dialogue TTS outperformed other formats in both ``perceived highest learning effect (self-perception)'' (47.3\%) and ``most enjoyable to learn from (preference)'' (66.9\%), and the deviations from a uniform distribution were highly significant in both cases. These are metrics of self-perception and preference, not evidence of objective learning outcomes.
\rev{The distinction is not a formality here, because the preference results and the outcome results point in different directions. Two thirds of learners named the dialogue format the most enjoyable to learn from and nearly half judged it the most effective, yet on the twelve primary outcomes the same session showed no advantage on perceived learning and was rated worse on audio quality. A format can be preferred while being no more effective on the measures available, and reporting preference alongside outcomes without marking the difference would invite exactly the conflation this study is not in a position to resolve. We therefore keep these measures in a separate section and exclude them from the outcome analysis. What they establish is that learners liked the dialogue format, which is a reason to consider it and not a demonstration that it works.}
Dialogue TTS also exceeded in-person sessions (35.4\%), instructor-voice video (27.7\%), and single TTS (22.3\%) for ``would like to experience again'' (50.5\%).

\subsection{Summary of Findings for Research Questions}

Table~\ref{tab:rq-summary} summarizes the validation results for the \rev{three} research questions.
\rev{For RQ1, none of the four core metrics differed between the three video formats, and every equivalence test placed the difference from instructor voice within the equivalence bounds, so TTS audio showed no detectable inferiority to a recorded human voice on these measures. For RQ2, the exploratory between-session comparison showed no general advantage for the dialogue session: three of twelve primary outcomes differed, with higher confidence in one's own understanding set against lower audio quality and higher extraneous load, and the remaining nine showing no difference. For RQ3, the generation logs yield a set of review criteria that the educators applied in practice, including one instance in which an internal contradiction in the generated material was caught before delivery.}

\begin{table*}[htbp]
\caption{Summary of Validation Results for Research Questions}
\label{tab:rq-summary}
\centering
\small
\begin{adjustbox}{max width=\textwidth}
\begin{tabular}{cp{4.2cm}p{5.8cm}p{3cm}}
\toprule
RQ & Question & Key results & Conclusion \\
\midrule
\multirow{3}{*}{RQ1} & \multirow{3}{4.2cm}{Does TTS audio substantially degrade the learning experience compared with instructor voice?} & Friedman test: $p \geq .142$ on all four metrics; Kendall's $W \leq .011$ & \multirow{3}{3cm}{No detectable degradation. Equivalence supported within the stated bounds} \\
& & TOST: all 8 comparisons within $\Delta=\pm0.5$; widest 90\% CI $[-0.212, +0.015]$ & \\
& & Free-text: consistency and intelligibility set against improvisation and expressiveness & \\
\midrule
\multirow{4}{*}{RQ2} & \multirow{4}{4.2cm}{How do ratings of the dialogue TTS lesson differ from those of the single-speaker TTS lesson?} & Confidence in own understanding higher: $r=+.130$, $q=.025$ & \multirow{4}{3cm}{Exploratory. A mixed pattern, not an advantage. Nine of twelve outcomes did not differ} \\
& & Audio quality lower: $r=-.182$, $q=.010$; extraneous load higher: $r=+.171$, $q=.010$ & \\
& & Perceived learning, attention, relevance, and all three engagement facets: no difference & \\
& & Free-text attributes the audio cost to voice drift between generation calls and to session design & \\
\midrule
\multirow{2}{*}{RQ3} & \multirow{2}{4.2cm}{What do educators review and revise in LLM-generated materials?} & Seven review criteria derived from the complete generation logs of both lessons & \multirow{2}{3cm}{Review is substantive, not nominal} \\
& & Includes a logical inconsistency in a worked example, corrected before delivery & \\
\bottomrule
\multicolumn{4}{l}{\footnotesize $r$: rank-biserial correlation; positive values favor the dialogue TTS session. $q$: BH-FDR corrected across the twelve outcomes.} \\
\multicolumn{4}{l}{\footnotesize RQ2 compares sessions whose respondents likely overlap without matching; its tests assume independence and are exploratory.} \\
\multicolumn{4}{l}{\footnotesize Preference measures are reported separately in Section~\ref{sec:preference}; they record what learners liked, not what they learned.} \\
\end{tabular}
\end{adjustbox}
\end{table*}

\subsection{Practical Implications}
\label{sec:implications}

\rev{Two decisions face an educator considering this system, and the results speak to them differently.}

\rev{The first is whether to use synthesized narration at all. Here the evidence is reasonably clear. Learners who had experienced both did not rate a synthesized lesson below one narrated by their teacher's recorded voice on any of the four metrics, and equivalence testing supported that reading rather than merely failing to reject a difference. Since recording narration is the step that makes producing lesson videos expensive, and since the alternative learners were actually offered was a recording rather than a live teacher, this is the finding with the most direct practical consequence.}

\rev{The second is whether to use two speakers rather than one, and here the evidence does not support a general recommendation. The dialogue format was not rated higher on perceived learning, attention, relevance, or any facet of engagement. What it was rated higher on was confidence in one's own understanding, and it was rated lower on audio quality and higher on extraneous load. An educator whose aim is that learners come away willing to engage with a topic may still find the format worth trying, particularly since two thirds of learners named the dialogue format the most enjoyable to learn from. An educator producing material where listening conditions are demanding, or where learners have difficulty with auditory comprehension, should prefer the single-speaker mode. What the results do not support is adopting the dialogue format on the expectation that it will improve understanding.}

\rev{The audio cost is also contingent rather than inherent. It arises because the synthesis model does not hold voice characteristics constant across separate generation calls, and the lesson is generated one slide at a time. A model that guarantees consistent voice characteristics would remove it, and until then, displaying speaker names or subtitles supplies speaker identity through a channel that does not depend on the audio.}

Regarding directions for system improvement, first, visual representation of speaker identity through subtitles, avatars, and speaker name overlays is considered effective for reducing extraneous load.
Second, \rev{prosody was the most frequent audio complaint in both conditions, so refinement of prosody control, or a model with stronger prosody in the language of instruction, would improve the experience of synthesized lessons generally rather than of the dialogue format specifically.}
Third, the scaffolding function can be further enhanced through adaptive adjustment of question strategies and novice utterance volume according to learner level and subject.

\subsection{\rev{Generalizability Beyond the Japanese Context}}
\label{sec:generalizability}

\rev{This study was conducted in one Japanese high school, in Japanese, within a curriculum slot for inquiry-based cross-disciplinary study that has no exact counterpart in every school system. Several of the findings are tied to that setting more closely than others, and the distinction is worth stating.}

\rev{The audio results are the most language-dependent. The most frequent complaint in both sessions concerned intonation, and one student identified a compound that the engine had read with the wrong pronunciation. Both are characteristic failure modes for Japanese speech synthesis rather than general ones: Japanese is a pitch-accent language in which accent placement distinguishes words, and its writing system admits multiple readings for the same characters, so the same underlying model can be substantially more or less accurate in Japanese than in a language where neither problem arises. The magnitude of the audio disadvantage we observed should therefore not be transferred to other languages without re-measurement, and the same applies to the equivalence result for RQ1, which holds for a synthesis quality that other languages may exceed or fall short of.}

\rev{The dialogue design is also language-dependent in a way that is easy to overlook. Japanese encodes the relation between speakers grammatically, through honorific and politeness forms that must be chosen for every utterance. An Expert--Novice exchange in Japanese therefore declares a status relation whether or not the designer intends it, and the personas carry social information that the same script in English would not. Whether the persona pairing we used would function equivalently in a language without this obligation is an open question.}

\rev{What is more likely to transfer is the workflow. The review criteria in Section}~\ref{sec:hitl}\rev{ concern completeness, technical viability, factual consistency, provenance, fit to delivery mode, fit to the learner, and continuity of the narration as speech; none of these depends on the language of instruction, and the failure they were responding to---a generated worked example that contradicted itself---is a property of how such models produce plausible text rather than of Japanese. The lower audio ratings of the dialogue session, to the extent that the free-text responses correctly attribute them to voice variation between generation calls, likewise reflect a structural property of generating two voices from a model whose output varies between calls, and should be expected wherever that property holds.}

\rev{Finally, the participants were first-year high school students in a single prefecture, following a curriculum designed and delivered by the authors. Learners who are older, who are studying vocationally, or who have less exposure to online video instruction may weigh consistency against expressiveness differently, and the population here reported using online video for study at a rate that is itself a characteristic of this cohort. Replication with other populations is required before the acceptability finding is treated as general.}

\subsection{Theoretical Contributions}

\rev{Three contributions follow from these results, stated at the strength the data support.}

\rev{First, for Cognitive Load Theory, the results identify where the cost of a multi-speaker narration actually falls. Adding a second speaker was accompanied by higher extraneous load and lower audio quality, but the free-text responses locate that cost in the instability of the synthesized voice between generation calls rather than in the dialogue structure itself, and complaints about prosody were equally frequent in both formats. The prediction that a second speaker imposes extraneous load is therefore not tested cleanly by this design; what is shown is that a synthesis engine whose voice characteristics drift across calls imposes such a cost, and that the cost is a property of the generation method rather than of dialogue as an instructional form.}

\rev{Second, for accounts of social cues in multimedia learning, two observations bear on the boundary of the effect. A recorded human voice was credited by learners with improvisation and responsiveness it did not possess, which suggests that the attribution follows from knowing the voice is human rather than from anything in the delivery. And within the dialogue condition, the items concerning the Novice persona did not form a separate dimension but loaded with items about how the material was presented, which is more consistent with the persona operating as a feature of presentation than as a distinct channel of social engagement. Both observations concern learners' perceptions and neither establishes an effect on learning.}

\rev{Third, the study contributes a documented account of what educator review consists of in an LLM-assisted production workflow. Because the generation records were retained in full, the review step could be characterized from evidence rather than asserted, yielding seven criteria including one---continuity of narration as continuous speech---that arises only when the output is heard rather than read. The record includes a self-contradictory worked example caught before delivery, which is the kind of error such a step is intended to catch and is not detectable by fluency alone.}

\subsection{Limitations and Future Directions}
\label{sec:limitations}

This study has several methodological limitations.
First, because the order of implementation of the three formats was fixed and the lesson content differed across sessions, the effects of lesson format cannot be completely separated from content differences. In particular, format differences in ``useful in the future'' are likely attributable to content differences.
Second, although the same cohort of students likely responded to both the single TTS questionnaire ($N=229$) and the dialogue TTS questionnaire ($N=206$), individual correspondence cannot be established due to anonymous collection, so the independence of observations is not guaranteed. Because we cannot control for dependence in this study, the repeated cross-sectional comparison is treated as an approximation, and the possibility of selection bias cannot be ruled out.
Third, only immediate subjective evaluation was conducted, and long-term effects such as knowledge retention, transfer, and sustained viewing behavior were not measured.
\rev{Fourth, the three sessions also differed in length: 20 min 17 s, 41 min 47 s, and 22 min 14 s for instructor voice, single TTS, and dialogue TTS respectively. The single TTS session was roughly twice as long as the other two, so fatigue and sustained attention differ across sessions in a way that is not separable from format. Session length was set by the curriculum slot rather than by the format under study, but the confound is real and applies to every between-session comparison reported here.}

\rev{Fifth, the study changed the baseline against which it was measured. The proportion of respondents reporting prior exposure to machine-read instructional material rose from 44.5\% to 69.9\% between the two questionnaires, because the first session was itself such an exposure for many students. Later sessions were therefore evaluated by learners for whom synthesized narration was no longer novel, and novelty and familiarity run in opposite directions across the sequence.}

\rev{Sixth, the questionnaire items were created for this study. Reliability and dimensionality are now reported in Section}~\ref{sec:psychometrics}\rev{, and the analyses use composites only where the examination supports them. Three limitations remain. Six constructs are measured by a single common item each, so their measurement error is unknown and no reliability can be claimed for them. The factor structure does not reproduce the theoretical divisions from which the items were drawn, so claims at the level of individual ARCS components are not supported by these data, and we have avoided them. And two items differ slightly in wording between the two questionnaires and were matched manually; this affects the audio-difficulty item, which carries part of the extraneous load result.}

\rev{Seventh, an earlier face-to-face session was excluded from all analyses because it was held in a different physical setting whose effect on the ratings could not be separated from the effect of lesson format, as set out in Section}~\ref{sec:design}\rev{. The comparisons reported here are therefore restricted to lessons delivered as video in the same classroom, which is the appropriate comparison for the question of whether synthesized narration is acceptable, but it means this study says nothing about how video lessons compare with live instruction.}

To overcome these limitations, the following are needed in future work: first, a randomized crossover design \rev{with matched content and session length} to control for order and content effects (addressing the first limitation); second, tracking changes for individual learners using anonymous IDs (addressing the second limitation); and third, measurement of knowledge retention through pre- and post-tests (addressing the third limitation). \rev{The third of these is the one that would most change what can be concluded. Every outcome in this study is a learner's report about their own experience, including the confidence result, and no independent measure of what was learned was collected. Until such a measure is available, the question of whether the dialogue format affects learning, as opposed to how learning feels, remains open. For the narrower question of whether synthesized narration costs learners anything, Deng et al.}\cite{ref:deng2026}\rev{ do supply a retention measure and find no effect of voice type, which is reassuring for RQ1; no comparable measure exists for the single-speaker versus dialogue contrast of RQ2.}

Future research directions building on this study include verifying the effects of adding subtitles and avatars, and extending the approach to diverse learner populations such as university students and vocational trainees. In addition, quantitative evaluation of instructor effort time, correction volume, and decision criteria at each stage of the human-in-the-loop process, and development of a framework for quantitatively evaluating the overall efficiency and quality of the proposed system's workflow, are important topics. Note that analysis with prior knowledge as a covariate was conducted supplementally in this study (Section~\ref{sec:olr}), but expanding covariate analysis to include a broader range of individual-difference variables such as learning styles and technology experience remains a future task.

\section{Conclusion}

This study proposed a semi-automated system for generating dialogue-based lessons using LLMs and TTS technology, and exploratorily examined its educational potential through a practical quasi-experiment with 245 first-year high school students.

On the system design side, we constructed an architecture comprising three elements: a human-in-the-loop workflow, TTS-optimized narration generation, and automatic generation of Expert-$\times$-Novice dialogue format inspired by cognitive apprenticeship theory.

\rev{On the empirical side, three findings emerged, and they are of unequal strength.}

\rev{The first concerns whether synthesized narration is acceptable at all. When the same learners retrospectively compared the three video lessons they had experienced, none of the four evaluation metrics distinguished the formats, and every equivalence test placed the difference between each TTS format and the recorded instructor voice inside the stated bounds. The free-text responses suggest the two are rated on different strengths that offset one another: synthesized speech was credited with never being unintelligible, never stumbling, and holding a constant volume, while the human voice was credited with improvisation and expressiveness. This is the strongest result reported here, although it rests on retrospective self-reports, and it is the one with the most direct practical consequence: it weakens the objection that replacing recorded narration with synthesis costs learners something they notice.}

\revb{The second concerns how the dialogue session was rated relative to the single-speaker session, a comparison that is exploratory because respondents likely overlap and could not be matched. Of twelve primary outcomes, nine did not differ, including perceived learning. The three that did form a mixed pattern rather than an advantage:}\rev{ learners in the dialogue session were more confident that they had understood the material, and rated the audio worse and the extraneous load higher. The free-text responses locate the audio cost not in the dialogue format itself but in the synthesis engine, whose voice characteristics drift between generation calls in a way that is tolerable with one speaker and disruptive with two, compounded by the fact that the dialogue session also carried a written exercise. Both causes are addressable, the first by models that hold voice characteristics constant across calls and the second by instructional design. Whether the higher confidence rating persists when the audio cost is removed is an open question that this study cannot answer, and the rating itself is a belief about one's own understanding rather than a measure of it.}

\rev{The third concerns the review step. Because the complete generation records for both lessons were retained, what the educators checked and changed could be described from the record rather than asserted. Seven criteria emerged, one of which---continuity of the narration as continuous speech---has no counterpart in text-based generation. The record also contains an instance in which the generated worked example of the framework being taught contradicted itself, declaring a candidate admissible while recording it as failing the admissibility condition, and was corrected before delivery. Review in this workflow is therefore substantive rather than nominal, though the record describes what review did address and cannot show what it missed.}

\rev{These findings are subject to the constraints of a practical quasi-experiment in which lesson format, content, order, and session length all vary together, in which every outcome is a learner's report about their own experience, and in which one session ran to twice the length of the others. They describe what happened in one curriculum at one school. Identifying the effect of lesson format itself would require a randomized crossover design with matched content and duration, individual linkage across sessions, and an independent measure of what was learned.}

\rev{Taken together, the contribution of this work is not a demonstration that dialogue-based synthesized lessons teach better. It is that synthesized narration proved acceptable to learners where a recorded human voice was the alternative, that the dialogue session was rated higher on confidence and lower on audio quality rather than better across the board, in an exploratory comparison that does not isolate the format, and that the review step in an LLM-assisted production workflow can be documented in enough detail to show what it catches. For educators deciding whether to adopt such a system, these are the terms on which the decision can be made.}

\section*{Acknowledgment}

We would like to express our sincere gratitude to the learners who participated in the experiment, and to all those who provided valuable guidance and feedback.
This work was supported by the JSPS Program for Forming Japan's Peak Research Universities (J-PEAKS), Grant Number JPJS00420240017.

\section*{Ethical Considerations}
This study deals only with anonymized educational evaluation data that do not identify individuals, and qualifies for exemption from ethical review under the regulations of the authors' institution.
In conducting the experiment, with the approval of the school principal, the study objectives were explained to the participants (high school students) and their guardians, consent was obtained, and questionnaires were collected anonymously.
Data were used solely for research purposes in anonymized form.

\section*{Disclosure of AI Use}
This study includes the generation of educational content using AI (LLMs) as its research subject. The accuracy of the generated content was verified by the authors. LLMs were also used for proofreading and translation of the manuscript.


\end{document}